\documentclass[aps, 10pt, prd,
               notitlepage, twocolumn, superscriptaddress,
               longbibliography,
               nofootinbib, floatfix]{revtex4-2}

\usepackage{amsmath}
\usepackage{amssymb}
\usepackage{amsfonts}
\usepackage[utf8]{inputenc}
\usepackage[T1]{fontenc}
\usepackage{mathrsfs}
\usepackage{anyfontsize}
\usepackage{microtype}
\usepackage{colortbl}
\usepackage{tabularx}
\usepackage{enumitem}

\usepackage[linktocpage,breaklinks]{hyperref}
\usepackage[dvipsnames]{xcolor}

\usepackage{tensor}
\usepackage{bm}

\usepackage{graphicx}
\usepackage{epsfig}
\usepackage{epstopdf}

\usepackage{natbib}
\usepackage{tcolorbox}
\usepackage[normalem]{ulem}

\hypersetup{colorlinks=true,
            citecolor=NavyBlue,
            linkcolor=NavyBlue,
            urlcolor=NavyBlue}

\usepackage{dsfont}
\usepackage{tcolorbox}
\usepackage{pgfgantt}
\usepackage{standalone}
\usepackage{tikz}
\usetikzlibrary{external}
\usetikzlibrary{arrows}
\usetikzlibrary{decorations.pathmorphing}
\usetikzlibrary{hobby}

\usepackage{floatrow}

\newcommand{\tend}{t_{\rm end}}
\newcommand{\rt}{r_{\ast}}
\newcommand{\rtl}{\tilde{r}}
\newcommand{\real}{\textrm{Re}\,}
\newcommand{\imag}{\textrm{Im}\,}

\newcommand{\cs}{c_\textrm{s}}
\newcommand{\css}{c_\textrm{s}^2}
\newcommand{\rh}{r_\textrm{h}}
\newcommand{\ps}{\,(+)}
\newcommand{\mn}{\,(-)}
\newcommand{\nn}{\nonumber \\}
\newcommand{\dd}{{\rm d}}
\newcommand{\ee}{{\rm e}}

\newcommand{\rr}{{\rm r}}
\newcommand{\ii}{{\rm i}}

\newcommand{\lm}{\ell m}

\newcommand{\tp}{t_{\rm peak}}

\newcommand{\pd}{\partial}

\newcommand{\dV}{{\rm d}^{4}x \, \sqrt{-g}}
\newcommand{\lamo}{\lambda_{\rm o}}
\newcommand{\lame}{\lambda_{\rm e}}
\newcommand{\lameo}{\lambda_{\,\rm e,\, o}}

\newcommand{\aei}{\affiliation{Max Planck Institute for Gravitational Physics (Albert Einstein Institute),
D-14476 Potsdam, Germany}}
\newcommand{\uiuc}{\affiliation{Department of Physics and Illinois Center for Advanced Studies of the Universe,\\
The Grainger College of Engineering, University of Illinois Urbana-Champaign, Urbana, Illinois 61801, USA}}
\newcommand{\um}{\affiliation{Departamento de F\'isica, Universidad de Murcia, Murcia, E-30100, Spain}}
\newcommand{\utub}{\affiliation{Theoretical Astrophysics, University of T\"ubingen,
Auf der Morgenstelle 10, D-72076 T\"ubingen, Germany}}
\newcommand{\uv}{\affiliation{Department of Physics, University of Virginia, Charlottesville, Virginia 22904, USA}}
\newcommand{\ethz}{\affiliation{Institut f\"ur Theoretische Physik, ETH Z\"urich, 8093 Z\"urich, Switzerland}}

\begin{document}

\title{Quasinormal modes and their excitation beyond general relativity. II:\\ isospectrality loss in gravitational waveforms}

\begin{abstract}
We continue our series of papers where we study the quasinormal modes, and
their excitation, of black holes in the simplest beyond general relativity
model in which first-principle calculations are tractable: a nonrotating black hole
in an effective-field-theory extension of general relativity with
cubic-in-curvature terms.
In this theory, the equivalence between the quasinormal mode spectra associated
with metric perturbations of polar and axial parities (``isospectrality'') of the
Schwarzschild black hole in general relativity no longer holds.
How does this loss of isospectrality translate into the time-domain ringdown
of gravitational waves? Given such a ringdown, can we identify the two
``fundamental quasinormal modes'' associated to the two metric-perturbation
parities?
We study these questions through a large suite of time-domain numerical
simulations, by a prescription on how to relate the gauge-invariant
master functions that describe metric perturbations of each parity with the
gravitational polarizations.
Under the assumptions made in our calculations, we find that it is in general
difficult to identify either of the two fundamental modes from the time series,
although finding evidence for a non-general-relativistic mode is possible sometimes.
We discuss our results in light of our assumptions and speculate about what may
occur when they are relaxed.
\end{abstract}

\author{Hector O. Silva}    \uiuc \aei
\author{Giovanni Tambalo}   \ethz
\author{Kostas Glampedakis} \um  \utub
\author{Kent Yagi}          \uv

\maketitle
\tableofcontents
\newpage

\section{Introduction}
\label{sec:intro}

The response of black holes to linear perturbations is known to be
partially described by a superposition of quasinormal modes, associated
with complex-valued frequencies~\cite{Vishveshwara:1970zz,Press:1971wr,Leaver:1986gd}.
The real and imaginary parts of these quasinormal frequencies dictate the
oscillation frequency and damping time of each mode, respectively.
The hole's response thus exhibits a characteristic ringdown.

In general relativity, the end state of the ringdown is a Kerr black hole,
whose mass $M$ and spin $a$ completely determine its quasinormal-mode
spectrum~\cite{Leaver:1985ax}.
This suggests the idea that the identification of two or more quasinormal
frequencies from gravitational-wave data may be used to infer the remnant's
mass and spin.
In analogy to using spectral lines to identify atomic and chemical elements,
this idea is called black hole spectroscopy~\cite{Detweiler:1980gk}; see
Ref.~\cite{Berti:2025hly} for a review.
The materialization of this ``black hole spectroscopy'' program is a key scientific
goal of current and future gravitational wave observatories.

Can black hole spectroscopy be used to test fundamental aspects of general relativity?
A notable property of the quasinormal mode spectrum of the static limit of the
Kerr solution is that the spectrum associated to the two ways metric
perturbations can transform under parity
transformation~\cite{Regge:1957td,Zerilli:1970se,Zerilli:1970wzz}, often called
``axial'' and ``polar,'' are the
same~\cite{Chandrasekhar:1975zza,Chandra:SchiffLectures,Chandrasekhar:1975nkd};
see also Ref.~\cite{Glampedakis:2017rar}.
The spectra are said to be isospectral.
In vacuum general relativity, this property persists beyond the Schwarzschild
solution, holding perturbatively in a small-spin
expansion~\cite{Franchini:2023xhd}.

\begin{figure}[t]
\centering
\includegraphics{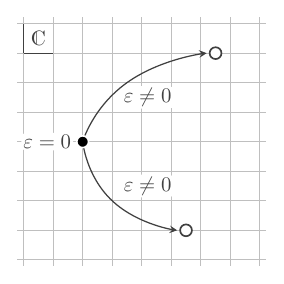}
\caption{Illustration of isospectrality breaking. Two degenerate quasinormal frequencies (black circle)
belonging to metric perturbations of opposite parities move apart in the complex frequency plane when the parameter $\varepsilon$
is tuned from zero (``general relativity'') to a nonzero value. Only one pair of frequencies is shown in the figure; however, this split also occurs, not necessarily in the same way, for all other frequencies. In our problem, $\varepsilon$ is
related to the lengthscale $l$ associated with higher-curvature terms in the EFT. Understanding how
isospectrality breaking impacts the black-hole ringdown is the main goal of this paper.}
\label{fig:iso}
\end{figure}

Isospectrality is, however, fragile and known to break in most extensions to
Einstein's theory; see Ref~\cite{Berti:2025hly}, Sec.~3. This is akin to the splitting of
degenerate spectral lines by an external electromagnetic field; see
Fig.~\ref{fig:iso}. (Theories that preserve isospectrality in some cases
are also known~\cite{Cano:2024wzo,Pope:2025jgz}.)

In our previous work~\cite{Silva:2024ffz}, we initiated a program to understand
ab initio the quasinormal modes and their excitation in the simplest
well-motivated beyond general relativity model in which first-principle
calculations are tractable: a nonrotating black hole in an
effective-field-theory (EFT) extension of general relativity with
cubic-in-curvature terms.
Among other results, we reported the first calculation of quasinormal-mode
excitation factors beyond general relativity and began exploring the
implications of isospectrality breaking in the ringdown of black holes in this
EFT.

The purpose of this paper is to continue that work. In particular, we turn from
frequency-domain to time-domain calculations. We study the linear response of
black holes in the EFT to external perturbations, and suggest how we can use
these results to obtain the gravitational-wave polarizations.
These results will be used to investigate the impact of the presence of a
doublet of fundamental quasinormal mode frequencies in the ringdown, and to
what extent we can infer their presence in the waveform.

This work is organized as follows.
In Sec.~\ref{sec:action}, we review the EFT of general relativity,
as well as the background black-hole spacetime we study.
In Sec.~\ref{sec:perturbations}, we give a concise account of the two master
equations that describe linear metric perturbations to this black hole and
discuss their properties.
Next, in Sec.~\ref{sec:simulations}, we carry out time-domain numerical
simulations of these master equations. We study how different the resulting
waveforms are from their general-relativistic limit and examine their spectral
content.
In Sec.~\ref{sec:implications}, we give a prescription for combining these
waveforms to obtain the gravitational wave polarizations in the EFT and
discuss what information can and cannot be extracted from them.
In Sec.~\ref{sec:conclusions}, we summarize our findings, discuss their limitations,
and entertain some conjectures.
The main text is supplemented by three appendices.
In Appendix~\ref{app:code_description}, we describe our numerical code,
and present convergence tests and error estimates.
In Appendix~\ref{app:schwarz}, we discuss how a wave equation with a variable
propagation speed can be mapped into a wave equation with a constant
propagation speed but with a modified effective potential.
In Appendix~\ref{app:qnm_fit_med_var}, we present additional
simulations that explore the initial-data parameter space.

We use geometrical units $c=G=1$ and the mostly plus metric signature.
We denote by $l$ the lengthscale associated to higher-curvature corrections to
general relativity, $l^4$ being the order of the corrections we study.
We work perturbatively to leading order in the dimensionless parameter
$\varepsilon = (l / M)^4$, where $M$ is the black hole's mass.

\section{Action and field equations} \label{sec:action}

We consider the following action
\begin{equation}
    S = \frac{1}{16 \pi} \int \dV~R
    + \frac{1}{16 \pi} \sum_{n \geqslant 2} {l^{2n - 2}} \, S^{(2n)},
    \label{eq:action_schematic}
\end{equation}
where $l$ is a lengthscale assumed to be small compared to the lengthscale
associated with a black hole of mass $M$, i.e., $M \gg l$, and $S^{(2n)}$ is
the action of the $n$th order curvature term which has $2n$ derivatives of the
metric.
We then call $S^{(2n)}$ a ``dimension-$2n$ operator.''
Notice that only even powers in $l$ are allowed by dimensional analysis.

As long as the EFT construction is built around vacuum general relativity and by performing
field redefinitions, one can show that no dimension-four operator exists.
The first nontrivial contribution occurs at dimension six and, at this order,
there are only two operators~\cite{Cano:2019ore,deRham:2020ejn}.
The resulting action is
\begin{align}
S &= \frac{1}{16 \pi} \int \dV \,
[R + {l^{4}}  \mathscr{L}],
\label{eq:s6_action_final_cano}
\end{align}
where
\begin{equation}
    \mathscr{L} = \lambda_{\rm e} \, R_{\mu\nu}{}^{\rho\sigma} R_{\rho\sigma}{}^{\delta\gamma}R_{\delta\gamma}{}^{\mu\nu}
    + \lambda_{\rm o} \, R_{\mu\nu}{}^{\rho\sigma} R_{\rho\sigma}{}^{\delta\gamma} \tilde{R}_{\delta\gamma}{}^{\mu\nu} \,.
    \label{eq:l6}
\end{equation}
Here, $\tilde{R}_{\mu\nu\rho\sigma} = (1/2) \, \epsilon_{\mu\nu}{}^{\alpha\beta} \, R_{\alpha\beta\rho\sigma}$,
$\epsilon_{\mu\nu\rho\sigma}$ is the totally antisymmetric Levi-Civita
tensor, and $\lameo$ are dimensionless constants associated to the even- (``e'')
and odd-parity (``o'') curvature terms.

The field equations of the theory, obtained by varying the action~\eqref{eq:s6_action_final_cano} with respect to $g^{\mu\nu}$, are:
\begin{equation}
    \mathscr{E}_{\alpha\beta} = G_{\alpha\beta}
    + l^{4} \mathscr{S}_{\alpha\beta} = 0,
    \label{eq:field_equations_schematic}
\end{equation}
where
\begin{subequations}
\label{eq:field_equations}
\begin{align}
    \mathscr{S}_{\alpha\beta}
    &=
    P_{(\alpha}{}^{\rho\sigma\gamma} R_{\beta)\rho\sigma\gamma}
    - \tfrac{1}{2} g_{\alpha\beta} \mathscr{L}
    + 2 \nabla^{\sigma} \nabla^{\rho} P_{(\alpha|\sigma|\beta)\rho} \,,
    \nonumber \\
    \label{eq:def_field_eqs}
    \\
    P_{\alpha\beta\mu\nu}
    &=
    3 \lame R_{\alpha\beta}{}^{\rho\sigma} \, R_{\rho\sigma\mu\nu}
    \nonumber \\
    &\quad + \tfrac{3}{2} \lamo (R_{\alpha\beta}{}^{\rho\sigma} \tilde{R}_{\rho\sigma\mu\nu}
    + R_{\alpha\beta}{}^{\rho\sigma} \tilde{R}_{\mu\nu\rho\sigma}).
    \label{eq:def_p_tensor}
\end{align}
\end{subequations}
As in Ref.~\cite{Silva:2024ffz}, we will only consider the even-parity
operator; we set $\lamo = 0$ and write $\lame = \lambda$. We assume $\lambda$ to be
positive, though $\lambda$ can have either
sign a priori~\cite{Goon:2016mil,Caron-Huot:2022ugt,Horowitz:2023xyl}.
We will work to leading order in $\lambda$, that is, to $\mathcal{O}(l^4)$.
Other aspects of the EFT in the context of gravitational-wave physics are
discussed, for instance, in
Refs.~\cite{Sennett:2019bpc,AccettulliHuber:2020dal,Silva:2022srr,Cano:2022wwo,Cayuso:2023xbc,Brandhuber:2024lgl,Melville:2024zjq,Maenaut:2024oci,Figueras:2024bba,Bernard:2025dyh}
and references therein.

In the presence of the dimension-six operators, the Schwarzschild solution is
no longer a solution in the EFT. Yet, an analytical black hole solution
that reduces to Schwarzschild when $\lambda = 0$ can be found perturbatively.
Its line element in Schwarzschild coordinates is
\begin{equation}
    \dd s^2 =
    - N^2 f \, \dd t^2
    + f^{-1} \, \dd r^2
    + r^2 \, \dd \theta^2
    + r^2 \sin^2\theta \, \dd\phi^2,
    \label{eq:line_element}
\end{equation}
where the metric functions $N$ and $f$ are
\begin{subequations} \label{eq:resum_gtt_grr}
    \begin{align}
    \label{eq:gtt_resum_final}
    N^2 f &=
    \left(1 - \frac{\rh}{r}\right)\left[ 1 - \varepsilon \, \left(
    \frac{5 M}{8 r}+\frac{5 M^2}{4 r^2}+\frac{5 M^3}{2 r^3}+\frac{5 M^4}{r^4}
    \right. \right.
    \nonumber \\
    &\quad \left.\left. +\frac{10 M^5}{r^5}+\frac{20 M^6}{r^6}\right)\right],
    \\
    \label{eq:resum_f}
    f^{-1} &= \left(1-\frac{\rh}{r}\right)^{-1}  \, \left[
    1 + \varepsilon
    \left( \frac{5 M}{8 r} + \frac{5 M^2}{4 r^2} + \frac{5 M^3}{2 r^3} + \frac{5 M^4}{r^4}
    \right.
    \right.
    \nn
    & \left.\left. \quad
    + \frac{10 M^5}{r^5} - \frac{196 M^6}{r^6}
    \right)\right].
    \end{align}
\end{subequations}
We defined the perturbative parameter
\begin{equation}
    \varepsilon = \lambda (l / M)^4,
\end{equation}
where $M$ is the hole's Arnowitt--Deser--Misner mass,
and $\rh$ is the location of the event horizon,
\begin{equation}
    \rh = 2 M ( 1 - 5 \, \varepsilon / 16 ).
    \label{eq:event_horizon}
\end{equation}
The black hole is of Petrov-type D~\cite{Silva:2024ffz}. All our calculations
are carried out to leading order in $\varepsilon$.

\section{Black hole perturbations}
\label{sec:perturbations}

\subsection{Review of the main equations}
\label{sec:review}

In Ref.~\cite{Silva:2024ffz}, we studied the linear gravitational perturbations
of the black hole solution~\eqref{eq:line_element} in the metric-based
formalism pioneered by Regge, Wheeler, and Zerilli~\cite{Regge:1957td,Zerilli:1970se,Zerilli:1970wzz},
following, in particular, the conventions from Martel and Poisson~\cite{Martel:2005ir}.
Our approach to this problem in the EFT differed in some aspects from
that of Refs.~\cite{deRham:2020ejn,Cano:2021myl}. We refer the reader to
Ref.~\cite{Silva:2024ffz}, Sec.~3 for details.

In brief, the problem reduces to studying two wave equations:
\begin{equation}
    [ -\,c^{-2}_{\rm s}(r) \partial_{tt} + \partial_{\rt \rt} - V^{(\pm)}_{\ell}(r) ] \,
    X^{(\pm)}_{\ell}(t,r) =
    0.
    \label{eq:eqs_rwz}
\end{equation}
We use the superscript $(\pm)$ to denote variables associated to metric
perturbations of polar ($+$) or axial ($-$) parity, and label them by their
multipolar index $\ell \geqslant 2$.
Metric perturbations of polar and axial parities are completely described by a single master function known
as the Zerilli--Moncrief (ZM) $X^{(+)}$ and Cunningham--Price--Moncrief (CPM) $X^{(-)}$ functions, respectively.
These are close relatives of the Zerilli~\cite{Zerilli:1970se} and
Regge--Wheeler functions~\cite{Regge:1957td}, respectively.
These two pairs of master functions satisfy the same \emph{homogeneous}
differential equations both in the EFT and in the limit of general relativity.

\begin{figure}[t]
\includegraphics{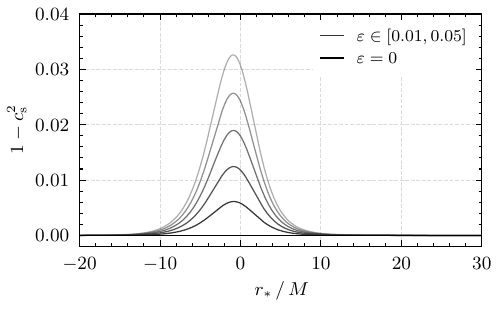}
\caption{Deviation from unity of the propagation speed squared of linear metric perturbations.
We vary the parameter $\varepsilon = \lambda \, l^4/M^4$ from zero (general relativity) to $0.05$
in increments of $0.01$. The maximum deviation from the speed of light happens at $r = 6 \rh / 5$,
near the hole's horizon $\rh$, and vanishes at the horizon and at spatial infinity.}
\label{fig:speed}
\end{figure}

In Eq.~\eqref{eq:eqs_rwz}, we also defined the tortoise coordinate $\rt$
\begin{equation}
\label{eq:def_tortoise}
    \dd \rt / \dd r = 1 / (Nf),
\end{equation}
which maps the domain $\rh \leqslant r < \infty$ to $-\infty < \rt < \infty$.
Details about the tortoise coordinate can be found in Ref.~\cite{Silva:2024ffz},
Appendix~D.
Also in Eq.~\eqref{eq:eqs_rwz}, $c_{\rm s}$ and $V_{\ell}^{\,(\pm)}$ are the
space-dependent propagation velocity of the perturbations and black-hole
effective potentials, respectively.
The former is given by~\cite{deRham:2020ejn}
\begin{equation}
    \css = 1 - 288 \, \varepsilon \left( 1 - \frac{\rh}{r} \right) \frac{M^5}{r^5}.
    \label{eq:sounds_speed}
\end{equation}
Note that $c_{\rm s}$ approaches unity at the event horizon and spatial infinity.
It also has a single extremum at $r = 6 \rh / 5$, at which the propagation speed deviates from unity by the amount
\begin{equation}
    ( 1 - \css) \big|_{r=\tfrac{6}{5}\rh} = \frac{3125}{162} \frac{M^5}{\rh^5} \, \varepsilon
             \simeq \frac{3125}{5184} \, \varepsilon \approx 0.603 \, \varepsilon,
\end{equation}
to leading order in $\varepsilon$.
The deviation can be super or subluminal depending on the sign of
$\varepsilon$~\cite{deRham:2020ejn}. We consider the latter case only.
Figure~\ref{fig:speed} shows the spatial profile of $\css$ in the range
$\varepsilon \in [0, 0.05]$ in increments of $0.01$.

\begin{figure*}[ht]
\includegraphics{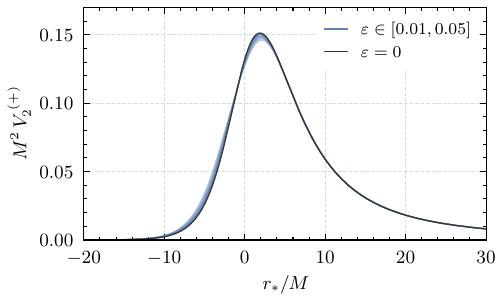}
\includegraphics{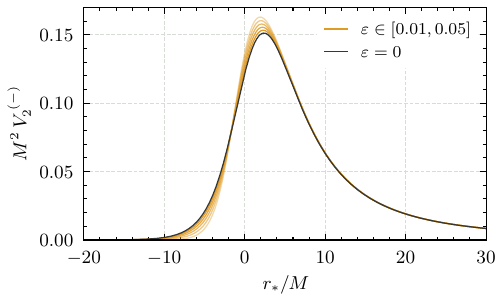}
\caption{
The effective potentials $V_{2}^{(\pm)}$ for perturbations of polar (left panel) and axial (right panel) parity. We vary the parameter $\varepsilon = \lambda \, l^4/M^4$ from zero (general relativity) to $0.05$
in increments of $0.01$.
Deviations from general relativity are mostly bound to the region between the event horizon, pushed to $\rt \to -\infty$, and the location of the potential peak.}
\label{fig:potentials}
\end{figure*}

We write the effective potential as
\begin{equation}
    V_{\ell}^{\,(\pm)} = \left(1 - \frac{\rh}{r} \right) \,
    \left[ \bar{V}_{\ell}^{\,(\pm)} + \varepsilon \, \delta V_{\ell}^{\,(\pm)} \right].
\end{equation}
The general-relativistic contributions to the potential are the
Zerilli~\cite{Zerilli:1970se} and Regge--Wheeler~\cite{Regge:1957td}
potentials,
\begin{subequations}
\label{eq:effective_potentials_gr}
\begin{align}
    \bar{V}_{\ell}^{\ps} &= \frac{1}{(r \Lambda_{\ell})^2} \left[
        2 \lambda_{\ell}^2 \left(\Lambda_{\ell} + 1\right)
        + \frac{18 M^2}{r^2} \left( \lambda_{\ell} + \frac{M}{r} \right)
    \right],
    \nonumber \\
    \label{eq:pot_zerilli}
    \\
    \bar{V}_{\ell}^{\mn} &= \frac{1}{r^2} \left[ \ell(\ell+1) - \frac{6M}{r} \right],
    \label{eq:pot_regge_wheeler}
\end{align}
\end{subequations}
respectively, where we defined:
\begin{equation}
    \lambda_{\ell} = (\ell + 2) (\ell - 1) / 2,
    \quad \textrm{and} \quad
    \Lambda_{\ell} = \lambda_{\ell} + 3M/r.
\label{eq:def_lambdas}
\end{equation}
The modifications to these potentials originating from the cubic-in-curvature terms have the form
\begin{subequations}
\label{eq:effective_potentials_eft_schematic}
\begin{align}
    \delta V_{\ell}^{\ps} &= \frac{1}{(r \Lambda_{\ell})^2} \, \sum_{k=1}^{10} v^{\ps}_{k \ell}(r) \, \left(\frac{M}{r}\right)^k,
    \\
    \delta V_{\ell}^{\mn} &= \frac{1}{r^2} \, \sum_{k=1}^{7} v^{\mn}_{k \ell} \, \left(\frac{M}{r}\right)^k.
\end{align}
\end{subequations}
The coefficients $v^{\ps}_{k \ell}$ depend on  $\Lambda_{\ell}$ for $k>4$,
hence their dependence on $r$.
In contrast, the coefficients $v^{\mn}_{k \ell}$ are independent of $r$ for all $k$.
The explicit form of $v^{\,(\pm)}_{k \ell}$ can be found in Ref.~\cite{Silva:2024ffz}, Appendix~E.\footnote{A cautionary note: the coefficients $v^{(+)}_{k \ell}$ for the corrections to the
Zerilli potential contain a typo in Ref.~\cite{Silva:2024ffz}; all coefficients
therein should be divided by four.}
Figure~\ref{fig:potentials} shows the spatial profile of $V_{2}^{(\pm)}$
in the range $\varepsilon \in [0, 0.05]$ in increments of 0.01.

The $\varepsilon$ corrections to the Zerilli and Regge--Wheeler potentials
break the equivalence between quasinormal-frequency spectrum of metric
perturbations of polar and axial parity that exists in general
relativity~\cite{Chandrasekhar:1975nkd,Chandra:SchiffLectures,Chandrasekhar:1975zza}.
The breakdown of isospectrality was first found through numerical calculations
in Refs.~\cite{deRham:2020ejn}, and its origin was discussed through the
analysis of a hierarchy of integral identities in Ref.~\cite{Silva:2024ffz}; see
also Refs.~\cite{Lenzi:2021njy,Lenzi:2025kqs}.
A particular EFT of general relativity involving quartic-in-curvature terms that
preserves isospectrality in the geometrical optics (eikonal) limit
was identified in Ref.~\cite{Cano:2024wzo} and
explored in Ref.~\cite{Cano:2025mht}.

We notice that Ref.~\cite{Cao:2025qws} studied Eq.~\eqref{eq:eqs_rwz} using a hyperboloidal
foliation of spacetime, whereas Ref.~\cite{Nakashi:2025fbr} considered the equivalent of Eq.~\eqref{eq:eqs_rwz}
in the quartic-in-curvature EFT of general relativity using double null coordinates.

\subsection{First-order form of the wave equation and initial data}
\label{sec:first_order_system}

We solve Eq.~\eqref{eq:eqs_rwz} using momentarily static initial data:
\begin{equation} \label{eq:id_static}
    \partial_{t} X_{\ell}^{(\pm)} |_{t=0}
    = 0,
\end{equation}
with compact support, localized far from the black hole.
In particular, we assume that $X_{\ell}^{(\pm)}$ has a Gaussian spatial profile
at $t = 0$,
\begin{equation} \label{eq:id}
    X_{\ell}^{(\pm)} |_{t=0}
    =
    A \, \ee^{-(\rt - r_{\ast}^{\rm med})^2 / (2 \sigma^2)},
\end{equation}
of width $\sigma$, amplitude $A$, and centered at
$r_{\ast}^{\rm med} \gg r_{\rm h}$. In all our simulations, we chose
\begin{equation} \label{eq:id_params}
    A = 1, \quad \sqrt{2} \, \sigma = 1.5M, \quad \textrm{and} \quad r_{\ast}^{\rm med} = 100M.
\end{equation}

We rewrite Eq.~\eqref{eq:eqs_rwz}, for each parity $(\pm)$ and
multipole $\ell$, as a system of first-order-in-time coupled partial
differential equations. For brevity, in the remainder of this subsection, we omit the scripts $(\pm)$
and $\ell$.
We perform the order reduction by introducing the auxiliary variables
\begin{equation}
\Pi = \pd_{t}X,
\qquad \textrm{and} \qquad
\Psi = \pd_{\rt} X.
\end{equation}
The system of equations that follows from Eq.~\eqref{eq:eqs_rwz} is then
\begin{subequations}\label{eq:first_order_pdes}
\begin{align}
    \pd_{t}\Pi  &= \css \, ( \pd_{\rt} \Psi - V X ),
    \\
    \pd_{t}\Psi &= \pd_{\rt} \Pi,
    \\
    \pd_{t}X    &= \Pi.
\end{align}
\end{subequations}

We evolve these equations forward in time using the method of lines.
The waveforms are extracted from the value of $X(t, \rt^{\rm ext})$ at an
extraction radius $\rt^{\rm ext} = 150M$.
Specifically, our code uses finite-difference
stencils that are fourth-order accurate in space and evolves the variables in time using the third-order accurate
Runge--Kutta scheme of Shu and Osher~\cite{Shu:1988JCoPh..77..439S}.
Details of our code are described in Appendix~\ref{app:code_description}, where
we also discuss the code's numerical convergence and give error estimates
on our waveforms.

\section{Numerical results} \label{sec:simulations}

We largely explored the quadrupole perturbations of axial and
polar parity.
We begin by revisiting the Vishveshwara's scattering experiments in general
relativity in Sec.~\ref{sec:simulations:gr}, and go beyond Einsteinian theory
in Sec.~\ref{sec:simulations:eft}.
In the latter case, we first study the effect of the variable propagation speed
on the evolution of the perturbations and how we can map this problem
into a wave equation with a constant propagation speed but with a modified effective potential
in Sec.~\ref{sec:simulations:eft:speed}.
We discuss how the waveforms in the EFT differ from those in general
relativity in Sec.~\ref{sec:simulations:eft:waveforms}, and study their
spectral content in Sec.~\ref{sec:simulations:eft:qnm_fit}.
We do so by extracting the fundamental quasinormal frequency through a fitting
procedure, which we compare against previous frequency domain calculations.\footnote{We remark that to our knowledge,
the first time-domain analysis on isospectrality breaking was done by Chaverra et al.~\cite{Chaverra:2016ttw}
in the context of linear perturbations of nonrotating black holes in nonlinear electromagnetism.
The isospectrality of the quasinormal mode spectra of axial and polar perturbations
of the Reissner--Nordstr\"om solution, shown by Chandrasekhar~\cite{Chandrasekhar:1980RSPSA.369..425C},
is generally broken in these nonlinear extensions to Maxwell's theory.}

\subsection{General relativity} \label{sec:simulations:gr}

Surprisingly, we have not found in the literature a scattering experiment
done with the Zerilli potential, with the exception of the somewhat
related work~\cite{GalvezGhersi:2019lag}.
It seems that since Vishveshwara~\cite{Vishveshwara:1970zz} the literature has
used the Regge--Wheeler potential exclusively. An ``exception'' is
Press, who performed time evolutions in the large-$\ell$ limit~\cite{Press:1971wr}.
In this limit, however, the Regge--Wheeler and Zerilli potentials become identical.

\begin{figure}[h]
\includegraphics{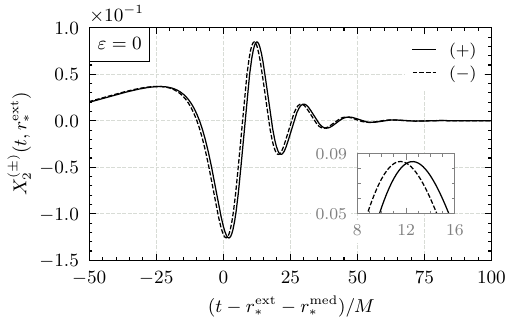}
\caption{Quadrupolar axial- and polar-parity waveforms in general relativity
obtained by evolving the same initial data. The inset zooms near the peak of
the waveforms and shows that they are offset by approximately $M$.}
\label{fig:rwz_gr_linear}
\end{figure}

For the sake of completeness, we present such a comparison in
Fig.~\ref{fig:rwz_gr_linear}.
We show the polar (solid line) and axial-parity (dashed line) waveforms
obtained by evolving the same initial data described in
Sec.~\ref{sec:first_order_system}.
The abscissa has been shifted by
the extraction radius, $r_{\ast}^{\rm ext} = 150M$, and the median position of
the initial data, $r_{\ast}^{\rm med} = 100M$.
The two waveforms are exquisitely similar to each other, except for a shift in
time of approximately $1M$.
This is most easily seen in the inset, wherein we zoom around a peak of the
waveforms. We attribute this difference to the different travel times required
by the same initial Gaussian wave packet to reach the peak of the effective
potential $V_{2}^{(\pm)}$.
These peaks differ by approximately $M$, occurring at larger values of $\rt$ for
the Regge--Wheeler potential than for the Zerilli potential. This causes the
axial-parity waveform to arrive at $\rt^{\rm ext}$ slightly earlier than its
polar-parity counterpart.
Shifting both waveforms in time according to their maximum value of
$|X_{2}^{(\pm)}|$ makes both virtually indistinguishable.
For this reason, wherever we show a waveform in general relativity in the next
sections, we have arbitrarily chosen to show the axial-parity waveform.
We note that Cunningham~et~al.~\cite{Cunningham:1978zfa,Cunningham:1979px} also
observed the similarity between waves of both parities in the perturbed
Oppenheimer--Snyder collapse~\cite{Oppenheimer:1939ue}.

\subsection{Beyond general relativity} \label{sec:simulations:eft}

After this brief discussion about waveforms in general relativity, we now
consider nonzero values of $\varepsilon$.

\subsubsection{Effects of the variable propagation speed} \label{sec:simulations:eft:speed}

\begin{figure}[t]
\includegraphics{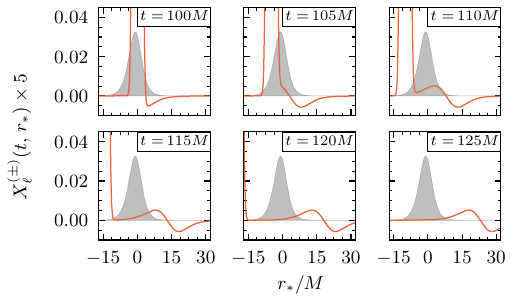}
\caption{Snapshots of the incident perturbation (solid line and rescaled by a factor of $5$) as it
propagates into the region of variable propagation speed. The filled gray curve represents $1 - \css$,
for $\varepsilon = 0.05$, as also shown in Fig.~\ref{fig:speed}. As the left-moving perturbation reaches the
region in which $\css$ deviates from one, part of it is reflected, as shown in the top panels. This reflected
pulse propagates rightward (see the bottom panels) and is measured by an observer at $\rt^{\rm ext}$ as the waveforms
in Fig.~\ref{fig:rwz_no_potential}.}
\label{fig:snapshots_variable_speed}
\end{figure}

We first study the effects of $\css$ on the waveforms.
To isolate the effects produced by this term, we begin by artificially setting
the potentials $V_{\ell}^{(\pm)}$ equal to zero in the wave equation~\eqref{eq:eqs_rwz} and
chose $\varepsilon = 0.05$ to maximize the effects of $\css$.
Because the potentials are zero, the resulting evolution is identical for
perturbations of axial and polar parity and is independent of $\ell$.
With this setup, we followed the evolution of the incident pulse as it entered
the region of space in which $\css$ deviates the most from one.
Figure~\ref{fig:snapshots_variable_speed} shows a sequence of snapshots of the
evolution. Specifically, we show the incident perturbation (solid line,
rescaled by a factor of 5) as it propagates into the region of variable speed.
This region is represented in the figure by the filled curve
superimposed in each panel; cf.~Fig.~\ref{fig:speed}. As the pulse
enters this region, part of it is reflected.
The bulk of the incident pulse then continues moving leftward, and a second
reflection occurs when the bulk of the incident pulse leaves the region in
which $\cs$ is not one. Eventually, the bulk of the pulse falls into the hole,
leaving behind a double-peaked wave that propagates to spatial infinity.

\begin{figure}[t]
\includegraphics{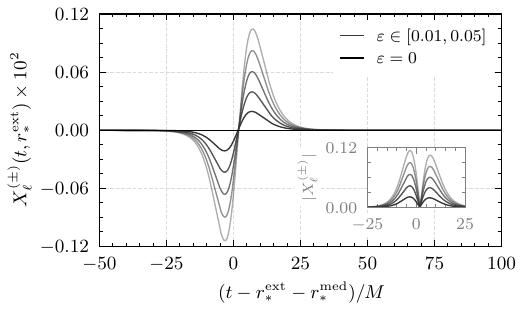}
\caption{Waveforms with varied $\varepsilon$ obtained by artificially setting the effective potential $V_{\ell}^{(\pm)}$ to zero
while maintaining a variable $\css$. The amplitude of the reflected wave grows with increasing $\varepsilon$,
i.e., with the increasing deviation from unit of $\css$; see Fig.~\ref{fig:speed}. In each waveform, the peak that arrives first has larger absolute value than the second, as shown in the inset. The asymmetry is due to the asymmetry of $\css$ with respect to the location of this maxima.}
\label{fig:rwz_no_potential}
\end{figure}

To understand how $\varepsilon$ influences the amplitude of the reflected wave,
we varied $\varepsilon$ from 0.01 to 0.05 in increments of 0.01.
Figure~\ref{fig:rwz_no_potential} shows the results of this exercise.
We show the waveforms extracted at a location $\rt^{\rm ext}$
far away from the hole. Unsurprisingly,
the amplitude of the reflected perturbation increases with
$\varepsilon$.
Through a curve fit, we find that maximum amplitude scales linearly with
$\varepsilon$, as expected, and is well approximated by
\begin{equation}
    \textrm{max}\,|X^{(\pm)}_{\ell}| \simeq 0.022460 \, \varepsilon,
\end{equation}
for the initial data and in the range of $\varepsilon$-values we studied.
This is approximately 5\% of the amplitude of the incident perturbation.
The structure of the scattered wave exhibits a small (but visible by eye)
asymmetry. This is most easily seen by plotting the absolute value of the
wave, as we do in the inset.
The asymmetry is a consequence of the asymmetry of $\css$ with respect to its
maximum; cf.~Fig.~\ref{fig:speed}.

\begin{figure}[t]
\includegraphics[width=\columnwidth]{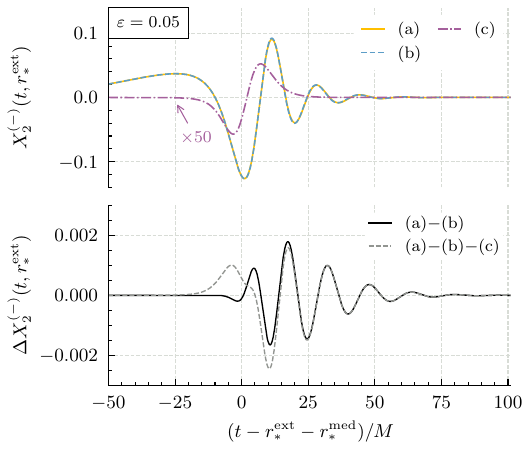}
\caption{Comparison between quadrupolar axial waveforms calculated when
    we set in the wave equation~\eqref{eq:eqs_rwz}: \ref{itm:case_a}~$\css \neq 1$ and $V^{(\pm)}_{\ell} \neq 0$,
   \ref{itm:case_b}~$\css = 1$ and $V^{(\pm)}_{\ell} \neq 0$,
and \ref{itm:case_c}~$\css \neq 1$ and $V^{(\pm)}_{\ell} = 0$. We consider $\varepsilon = 0.05$ in all cases.
In the top panel, we show the three waveforms.
We see that the effects of the variable propagation speed on the waveform are
subdominant relative to the presence of the effective potential; compare the
waveforms~\ref{itm:case_a} and~\ref{itm:case_b}, and their amplitude relative
to the waveform~\ref{itm:case_c}, which we rescaled by a factor of 50.
In the bottom panel, we subtract from the complete solution of Eq.~\eqref{eq:eqs_rwz}, the result
of~\ref{itm:case_b} (solid line) and the sum of cases~\ref{itm:case_b} and~\ref{itm:case_c} (dashed line).
}
\label{fig:speed_effects}
\end{figure}

How much does the variable propagation speed contribute to the final waveform
when we include the effective potential in our simulations?
To answer this question, we performed a second suite of simulations, evolving
both axial and polar-parity quadrupolar $\ell = 2$ perturbations where we
varied $\varepsilon$ from 0.01 to 0.05 in increments of 0.01 and artificially
set $\css = 1$, but including $V_{\ell}^{(\pm)}$.
Next, we ran a third set of simulations, but now including the effect of
$\varepsilon$ on both $\css$ and in the potentials; that is, we evolved the complete
Eq.~\eqref{eq:eqs_rwz}.
Together with the simulations discussed in
Figs.~\ref{fig:snapshots_variable_speed} and~\ref{fig:rwz_no_potential}, we
thus have three sets of simulations that we label as cases \ref{itm:case_a}, \ref{itm:case_b}, and \ref{itm:case_c}, as
summarized below:
\begin{enumerate}[label=(\alph*), wide, labelwidth=!, labelindent=0pt, leftmargin=*]
    \item \label{itm:case_a} $\css \neq 1$ and $V^{(\pm)}_{\ell} \neq 0$,
    \item \label{itm:case_b} $\css = 1$ and $V^{(\pm)}_{\ell} \neq 0$,
    \item \label{itm:case_c} $\css \neq 1$ and $V^{(\pm)}_{\ell} = 0$.
\end{enumerate}

In Fig.~\ref{fig:speed_effects}, we show an illustrative waveform of each case,
namely, for quadrupolar axial perturbations and $\varepsilon = 0.05$. In the top panel, we
plot the three waveforms for case~\ref{itm:case_a} (solid line),~\ref{itm:case_b} (dashed line)
and~\ref{itm:case_c} (dot-dashed line, rescaled by a factor of $50$).
We find that the contribution of $\css$ is small [compare the waveforms for
cases~\ref{itm:case_a} and~\ref{itm:case_b}] and subdominant relative to the contribution of the effective
potential [compare the waveforms for case~\ref{itm:case_c} against those of cases~\ref{itm:case_a} and~\ref{itm:case_b}.]
In the bottom panel, we show the difference between the ``complete'' waveform~\ref{itm:case_a} to case~\ref{itm:case_b}
(solid line) and when we subtract from \ref{itm:case_a} the sum of waveforms of cases \ref{itm:case_b} and \ref{itm:case_c} (dashed line).
We see that the differences are ${\mathcal O}(10^{-3})$, comparable to the order of magnitude of the waveform corresponding
to case~\ref{itm:case_c}, shown in Fig.~\ref{fig:rwz_no_potential}.
These observations are also shared by the polar-parity perturbations.

Before closing this section, we remark that the variable-speed wave
equation~\eqref{eq:eqs_rwz} can be recast as a constant-speed wave equation~\cite{Trachanas},
\begin{equation} \label{eq:eqs_rwz_schwarz}
    [ -\partial_{tt} + \partial_{\rtl\rtl} - W^{(\pm)}_{\ell}(r) ] \,
    Y^{(\pm)}_{\ell}(t,r) =
    0,
\end{equation}
for new master functions $Y^{(\pm)}_{\ell}$, coordinate $\rtl$, and
effective potential
\begin{equation}
    W^{(\pm)}_{\ell} = \css \, V^{(\pm)}_{\ell} - \{\rt, \rtl\},
\end{equation}
where $\{\rt, \rtl\}$ is the Schwarzian derivative defined as
\begin{equation} \label{eq:def_schwarz_deriv}
    \{x, y\} =
    \frac{1}{2} \frac{\dd}{\dd y}\left(\frac{\ddot x}{\dot x}\right)
    -
    \frac{1}{4} \left(\frac{\ddot x}{\dot x}\right)^2,
    \quad
    \dot{x} = \dd x / \dd y.
\end{equation}
We present the derivation of Eq.~\eqref{eq:eqs_rwz_schwarz}, discuss the
properties of $W^{(\pm)}_{\ell}$, and report the result of an illustrative
simulation (as well as of an unsolved puzzle) in Appendix~\ref{app:schwarz}.
Hereafter, we use Eq.~\eqref{eq:eqs_rwz} for our
simulations.

\subsubsection{Waveform comparison} \label{sec:simulations:eft:waveforms}

We now focus exclusively on the waveforms in case~\ref{itm:case_a}, the
solutions to Eq.~\eqref{eq:eqs_rwz}, and compare them against their
general-relativistic counterparts. We begin by considering $\varepsilon =
0.05$, the largest value of $\varepsilon$ we considered.

\begin{figure}[b]
\includegraphics{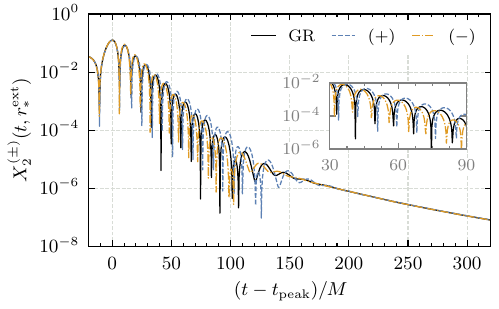}
\caption{Comparison between waveforms in general relativity and in the effective field theory.
The solid line corresponds to the CPM waveform in general relativity (``GR''), while the
dashed and dot-dashed lines represent the ZM, ``$(+)$,'' and CPM, ``$(-)$,'' waveforms when $\varepsilon = 0.05$.
The waveforms were shifted in time such that their peaks occur at zero. The inset shows the waveforms
in a time interval in which the signal is dominated by the quasinormal frequencies of the black hole.}
\label{fig:axipolar}
\end{figure}

\begin{figure*}[t]
\includegraphics{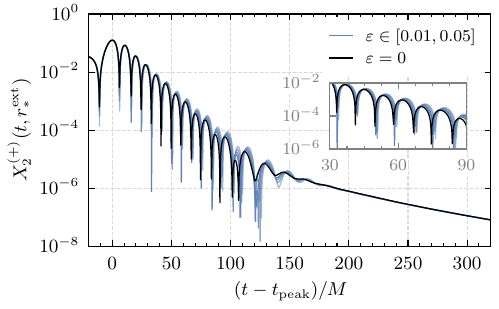}
\quad \quad
\includegraphics{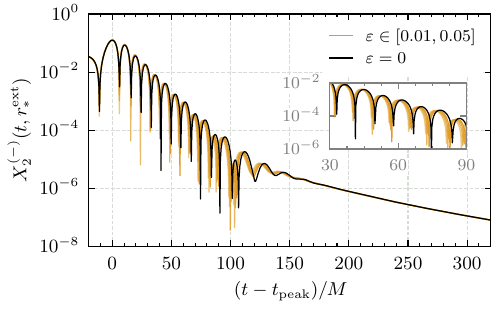}
\caption{Comparison between waveforms in general relativity and in the effective field theory for
various values of $\varepsilon$.
The solid line corresponds to the parity odd waveform in general relativity, while the
{blue and orange}
lines represent the even-parity (left panel) and odd-parity (right panel) waveforms for $\varepsilon \in [0, 0.05]$
in increments of 0.01. As in Fig.~\ref{fig:axipolar}, the inset shows the waveforms
in a time interval in which the signal is dominated by the quasinormal frequencies of the black hole.}
\label{fig:rwz_eft_log}
\end{figure*}

In Fig.~\ref{fig:axipolar}, we show the quadrupolar waveforms $X_{2}^{(\pm)}$ in
logarithmic scale. We shifted the abscissa such that the time instant $\tp$ at which the
maximum (absolute) value of the waveform occurs is zero.
The three curves represent the cases of general relativity (solid line), and
polar (dashed line) and axial (dot-dashed line) perturbations in the effective
field theory with $\varepsilon = 0.05$.
After doing this alignment in time, we see that all three waveforms are very
similar initially. They start to become visually different as we enter the
``ringdown'' stage that is dominated by the quasinormal frequencies of the
black hole. After these frequencies have been sufficiently damped, the signal
is dominated by a power-law tail that is identical for the three waveforms.
Because the damping of the quasinormal frequencies of polar and axial
perturbations is shifted differently by $\varepsilon$, the power-law tail of
the respective waveforms become prominent at different times.
Specifically, it occurs later for the polar perturbations in comparison
to the axial perturbations.
In the inset, we zoom into the waveforms' ringdown.
Qualitatively, we see two trends.
First, the oscillation frequency increases (decreases) for the axial
(polar) waveform with respect to general relativity.
Second, the damping of the oscillations is faster (slower) for the axial
(polar) waveform with respect to general relativity.
These are the expected outcomes from the frequency-domain calculations of the
fundamental quasinormal frequency carried out in Ref.~\cite{Silva:2024ffz}.
We will do a quantitative comparison between the spectral content of our waveforms
against these frequency-domain results in
Sec.~\ref{sec:simulations:eft:qnm_fit}.

Figure~\ref{fig:rwz_eft_log} is similar to Fig.~\ref{fig:axipolar} except that we
include intermediate values of $\varepsilon$ from zero up to $0.05$ in steps of $0.01$.
To reduce visual clutter, we separate the even- and odd-parity waveforms into
the left and right panels, respectively.
As expected, waveforms with smaller values of $\varepsilon$ are more similar to
the general-relativistic waveform, while the differences share the general
trends shown in Fig.~\ref{fig:axipolar} for $\varepsilon = 0.05$.
Finally, we see after $t$ approximately $180M$ after the peak, all waveforms
exhibit the same power-law tail behavior.

\subsubsection{Spectral content of the waveforms} \label{sec:simulations:eft:qnm_fit}

Having gained some understanding of the different ingredients that make up our
waveforms, we now study them in more detail. In particular, we address
two items.
First, we study their spectral content. In particular, we
extract the quasinormal frequencies in the signals and compare them against
our previous frequency-domain calculations~\cite{Silva:2024ffz}.
Second, we provide a quantitative measure of how much the waveforms
differ from those in the limit of general relativity. We address these
questions at the level of the metric-perturbation master functions $X^{(\pm)}$
in this section.
In Sec.~\ref{sec:implications}, we will discuss the observational prospects of the
results obtained herein by relating $X^{(\pm)}$ to the gravitational wave
polarizations.

Let us begin with the quasinormal frequencies. Here, we follow
Ref.~\cite{Baibhav:2023clw}; see also
Refs.~\cite{Nee:2023osy,Thomopoulos:2025nuf}.
A comprehensive review of quasinormal-frequency extraction strategies can be
found in Ref.~\cite{Berti:2025hly}.
We use as our fitting model for the $(\ell,m)$ multipole of the waveform a
linear superposition of damped sinusoids:
\begin{equation} \label{eq:qnm_fit}
    Q_{N} = \sum_{n = 0}^{N} A_{n} \, \ee^{- \omega_{\ii, n} (t - \tp)}
    \,
    \sin[\omega_{\rr, n}(t - \tp) + \phi_{n}],
\end{equation}
where $N$ denotes the number $n$ of overtones
in the model, $A_n$ their amplitude, $\omega_{\,\ii, n} = \imag \omega_{\ell n}$,
$\omega_{\,\rr, n} = \real \omega_{\ell n}$, and $\phi_{n}$ a phase.
The fit is applied to our numerical data in a time window
\begin{equation}
    t \in [t_0, \tend],
\end{equation}
where $t_0$ and $\tend$ are measured with respect to $\tp$.
Based on the conclusions of Refs.~\cite{Baibhav:2023clw,Nee:2023osy}, we focus
on extracting the fundamental quasinormal frequency only; i.e., we use $N=0$.
For this reason, we omit the subscript $n$ hereafter.

\begin{figure}[t]
\includegraphics{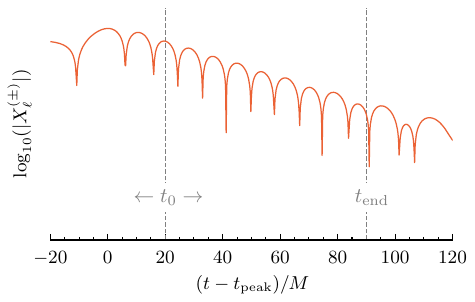}
\caption{Illustration of the variable time window in which we perform our fits.
We vary the initial time $t_0$ in the interval $t_0 \in [0, 30]M$, measured with respect to the
time instant in which the waveform has its largest value, $\tp$. The end
time is fixed to $\tend = 90 M$.}
\label{fig:illustration}
\end{figure}

To perform our fits, we keep $\tend = 90M$ fixed, while allowing $t_0$ to vary
in the interval $t_0 \in [0, 30]\,M$ in steps of $1M$; see
Fig.~\ref{fig:illustration} for an illustration. We found that our results are
insensitive to shifting $\tp$ up to $120M$.
For each time window, we fit Eq.~\eqref{eq:qnm_fit} against the numerical
data using the \texttt{curve\_fit} function from SciPy~\cite{2020SciPy}
to perform a least-square fitting. We use the ``Trust Region Reflective''
algorithm and set the bounds
\begin{equation*}
    A \in [0, 1], \,\,\,
    \phi \in [0, 2\pi), \,\,\,
    \omega_{\rr} \in [0, 2 \, \omega^{\rm GR}_{\rr}],\,\,\,
    \omega_{\ii} \in [0, 2 \, \omega^{\rm GR}_{\ii}],
\end{equation*}
on the free parameters of $Q_0$.
Here, $\omega^{\rm GR}_{\rm r,\,i}$ are the real and imaginary parts
of the fundamental quasinormal frequency of a Schwarzschild black hole~\cite{Chandrasekhar:1975zza}.
For $\ell = 2$, they are
\begin{equation} \label{eq:qnm_20}
    M \omega^{\rm GR}_{\rm r} = 0.373671,
    \quad \textrm{and} \quad
    M \omega^{\rm GR}_{\rm i} = - 0.0889623.
\end{equation}
We initialize the fitting algorithm by giving each free model parameter a random
initial guess, by drawing a sample from a uniform probability distribution
within the bounds on our parameters~\cite{Nee:2023osy}.

We repeat this fitting process 100 times for each value of $t_0$ to allow the
algorithm to converge to the true minima.
We then select from these 100 fits the tuple of best-fit parameters that gives
the smallest mismatch ${\cal M}$ between model and numerical data.
Consider two signals $h_{1,2}(t)$.
We define their inner product in the time interval $t \in [t_0,\,\tend]$ as
\begin{equation} \label{eq:def_inner_product}
    \langle h_1 | h_2 \rangle = \int_{t_0}^{\tend} \dd t \, h_1(t) \, h^{\ast}_2(t),
\end{equation}
where the asterisk denotes complex conjugation.
For our purposes the complex conjugation in Eq.~\eqref{eq:def_inner_product} is
inconsequential because our waveforms are real valued.
We then define the mismatch ${\mathcal M}$ as,
\begin{equation} \label{eq:def_mismatch}
    {\mathcal M} = 1 -
    \frac{\langle h_1 | h_2 \rangle}{\sqrt{  \langle h_1 | h_1 \rangle \, \langle h_2 | h_2 \rangle}}.
\end{equation}
Note that ${\mathcal M} = 0$ if the waveforms are identical, $h_1 = h_2$.
However, one must keep in mind that a small mismatch does not necessarily
imply a good fit to the data. For example, if $h_1$ and $h_2$ differ by
a nonzero multiplicative factor, their mismatch will be zero. Moreover, a small mismatch
does not exclude the possibility that some parameters in the fitting model are
overfitting the data; see Refs.~\cite{Baibhav:2023clw,Nee:2023osy} for a
discussion in the context of linear perturbation theory.

\begin{figure}[t]
\includegraphics[width=\columnwidth]{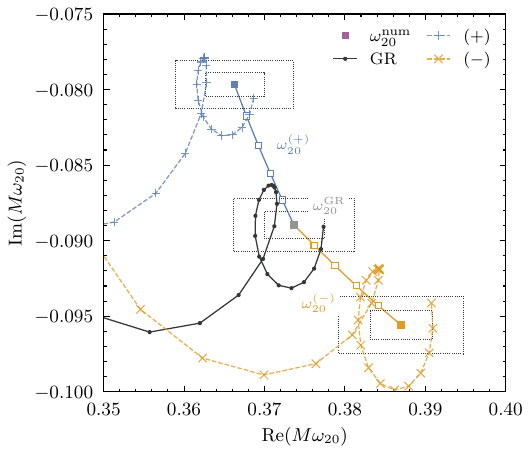}
\caption{Trajectories in the complex plane for the best-fit fundamental quasinormal frequencies $\omega_{20}$
as $t_0$ varies from zero to $30M$. The solid squares represent the frequencies calculated by us in Ref.~\cite{Silva:2024ffz}. The solid square at the center of the figure indicates $\omega_{20}$ of a Schwarzschild black hole,~\eqref{eq:qnm_20}, and that has the same value for axial and polar perturbations. This degeneracy is broken by nonzero values of $\varepsilon$. As
we increase this parameter, the polar (axial) quasinormal frequency moves upward (downward) in the complex frequency plane, as
indicated by the solid lines; intermediate values of the frequencies for $\varepsilon = 0.01$, 0.02, 0.03 and 0.04 are indicated
by empty squares.
The two curves terminate at $\varepsilon = 0.05$, and we indicate the frequency
values, cf.~Eq.~\eqref{eq:qnm_20_eft}, with the two additional solid squares.
The dotted-lined rectangles represent $\pm 1\%$ and $\pm2\%$ away from these
frequencies.
We see that as $t_0$ approaches $30M$, the best-fit frequencies approach their expected values determined
by the frequency-domain calculation of Ref.~\cite{Silva:2024ffz}.
}
\label{fig:l2n0_fits}
\end{figure}

Figure~\ref{fig:l2n0_fits} shows our results from applying this fitting process
to $X^{(\pm)}_{2}$. We show the trajectories in the complex
plane of the best-fit fundamental quasinormal frequency as we vary $t_0 = t - \tp$ from
zero to $30M$. The plot is somewhat busy, so let us focus for the moment on
the solid line labeled ``GR,'' for ``general relativity.''
Each circle along this curve represents a best-fit value of $\omega_{20}$ for a given value of $t_0$.
As $t_0$ increases, the trajectory moves from left to right, looping around
$\textrm{Re}(M\omega_{20}) \approx 0.37$. The trajectory then approaches the true
value~\eqref{eq:qnm_20}, indicated by a filled square, as $t_0$ approaches
$30M$. The dotted-lined rectangles represent values $\pm1\%$ and $\pm2\%$ away from~\eqref{eq:qnm_20}.

We now move away from the general-relativity curve by increasing $\varepsilon$.
A nonzero value of $\varepsilon$ breaks the degeneracy between $\omega^{(+)}$
and $\omega^{(-)}$ that exists in general relativity. Figure~\ref{fig:l2n0_fits} shows this
through the curves that branch up and downward from the
Schwarzschild frequency. The curves terminate at $\varepsilon = 0.05$, for which
\begin{subequations} \label{eq:qnm_20_eft}
\begin{align}
    M \omega_{20}^{(+)} &= 0.366287 - 0.0796875 \ii,
    \\
    M \omega_{20}^{(-)} &= 0.386990 - 0.0955811 \ii.
\end{align}
\end{subequations}
Quasinormal frequencies corresponding to the intermediate values of $\varepsilon = 0.01$,
0.02, 0.03, and 0.04 are indicated by the open squares.
For ``internal consistency,'' these values are quoted from our previous
work~\cite{Silva:2024ffz}; they agree to a few percent with the earlier works
by de Rham~et~al.~\cite{deRham:2020ejn} and Cano et~al.~\cite{Cano:2021myl}.
See Ref.~\cite{Silva:2024ffz}, Fig.~3, in particular.
We applied the fitting procedure to the waveforms for $\varepsilon = 0.05$.
The best-fit values for $\omega_{20}$ are shown by the dashed lines.
Different markers distinguish the polar and axial frequencies as described in
the figure's legend. The trajectories are similar to the case of general
relativity.
Satisfyingly, we were able to extract from our waveforms the values in
Eq.~\eqref{eq:qnm_20_eft} to 2\% precision when $t_0 = 30M$.

\begin{figure}[t]
\includegraphics{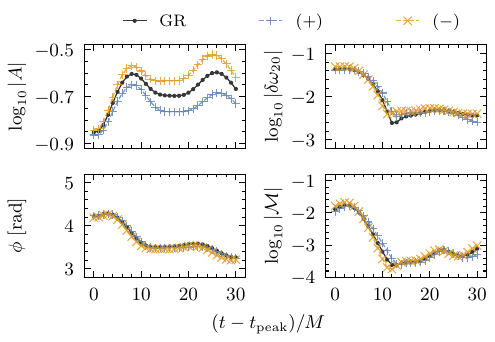}
\caption{Dependence of the mismatch and the best-fit parameters in our fitting model~\eqref{eq:qnm_fit} as a
function of $t_0$ for the three waveforms shown in Fig.~\ref{fig:axipolar}.
We see that the absolute errors $\delta \omega_{20}$ between the extracted
frequencies and their reference values become nearly constant for $t_0
\gtrsim 10 M$, and are smaller than 1\%.}
\label{fig:l2n0_fits_other_parameters}
\end{figure}

We note the potential degeneracy between general relativity and the effective
field theory depending on the time window used.
Figure~\ref{fig:l2n0_fits} shows that the general-relativity curve intersects
the nonzero-$\varepsilon$ curve. This occurs for $t_0 \approx 29M$. Moreover,
we see that depending on the value of $t_0$, the nonzero-$\varepsilon$ curves
for $\varepsilon = 0.05$ can intersect the exact values of $\omega_{20}$ for
\emph{other} values of $\varepsilon$.

In Fig.~\ref{fig:l2n0_fits_other_parameters}, we show the mismatch ${\cal
M}$ and best-fit values of the parameters in our model $Q_0$ as a function of
$t_0$ for the three waveforms we are analyzing.
Specifically, we show the best-fit amplitude $A$ (top left panel), absolute
error in the quasinormal frequency $\delta \omega_{20}$ (top right panel),
phase $\phi$ (bottom left panel), and mismatch ${\cal M}$ (bottom right panel).
Explicitly,
\begin{equation}
    \delta \omega_{20} = |\omega_{20}^{\rm best-fit} - \omega_{20}^{\rm num}|,
\end{equation}
where $\omega_{20}^{\rm num}$ is either
numerical values \eqref{eq:qnm_20} or~\eqref{eq:qnm_20_eft}, depending on the case.
The curves are similar, both in their qualitative behavior and in
magnitude. A notable difference is in the amplitude $A$, which increases (decreases)
with respect to the general-relativity curve for the axial (polar) waveform starting around $t_0 \gtrsim 6\,M$.
We attribute this behavior to the different damping times of the modes.
Lastly, we see that the mismatch stabilizes for $t_0 \gtrsim 10\,M$, although
it slowly increases around $30M$ after $\tp$.

\begin{figure}[t]
\includegraphics{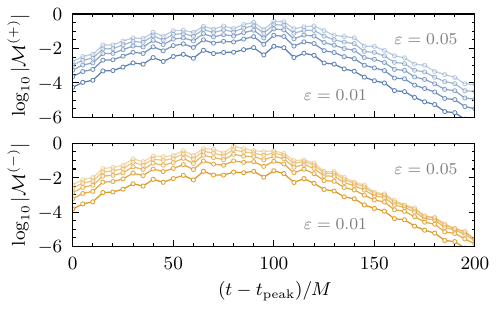}
\caption{Mismatch between polar (top panel) and axial (bottom panel) waveforms
with respect to general relativity as a function of beginning of the time window $t_0$.
The curves are similar, irrespective of the value of $\varepsilon$.
However, unsurprisingly, they shift toward larger values of ${\cal M}$ as $\varepsilon$ increases.
Focusing on the dependence on $t_0$, we see that ${\cal M}$ is largest when the
time interval contains the ringdown. As $t_0 \gtrsim 150M$, the time
window contains mostly the power-law tail and the mismatches rapidly decrease; see~Fig.~\ref{fig:rwz_eft_log}.}
\label{fig:mismatches}
\end{figure}

We also calculated the mismatch~\eqref{eq:def_mismatch} between our
waveforms in the EFT with respect to general relativity for
different values of $\varepsilon$ as a function of $t_0$.
To perform this calculation, we first aligned our nonzero-$\varepsilon$
waveforms with the peak of the general relativistic one.
Then we evaluated the inner product~\eqref{eq:def_inner_product}, varying $t_0
\in [-20, 200]M$ in increments of $5M$ while keeping $\tend = 230M$ fixed.
We show the results of this exercise in Fig.~\ref{fig:mismatches}. The behavior of the curves
as a function of $t_0$ is qualitatively the same for different values of $\varepsilon$.
Increasing $\varepsilon$ causes the mismatches to increase, as expected~\cite{Cao:2025qws}.
At fixed value of $\varepsilon$,
we see that the mismatches increase with increasing $t_0$; i.e., by having
mostly the ``ringdown-plus-tail'' part of the waveforms in the inner product~\eqref{eq:def_inner_product}.
Increasing $t_0$ further makes the inner product involve mostly the tail portion
of the signals and the mismatches decrease.

In Appendix~\ref{app:qnm_fit_med_var}, we show the outcome of performing this
fitting exercise to a small suite of simulations where the median $\rt^{\rm med}$
of our initial data is shifted.
The upshot of our discussion therein is that we are able to extract more
accurately the quasinormal frequencies the closer we make $\rt^{\rm med}$ to
the peak of effective potential.

\section{Observational implications} \label{sec:implications}

So far we have studied the properties of the waveforms $X^{(\pm)}$.
However, to make connection with gravitational-wave observables, we have
established a relation between the ZM and
CPM functions with the gravitational wave
polarizations $h_{+,\times}$.
In particular, we are interested in examining what happens when the even-
and odd-parity ringdown waveforms, each dominated by their own respective
fundamental quasinormal frequencies $\omega^{(+)}_{20}$ and $\omega^{(-)}_{20}$,
respectively, are combined to produce $h_{+,\times}$.
How does, for instance, $h_{+}$ behave, in comparison to general relativity, when we
have two distinct fundamental quasinormal frequencies $\omega^{(\pm)}_{20}$?
Are we able to extract the frequencies $\omega^{(\pm)}_{20}$ from this $h_{+}$
as we accomplished in Sec.~\ref{sec:simulations:eft:waveforms} from $X^{(\pm)}$?
We investigate these questions next.

\subsection{Gravitational wave polarizations}

In the context of general relativity, as discussed in Ref.~\cite{Martel:2005ir},
the gravitational wave polarization  $h_+$ and $h_\times$ at future null infinity
are related to the ZM and CPM master functions as
\begingroup
\allowdisplaybreaks
\begin{subequations} \label{eq:polarizations}
\begin{align}
    r \, h_{+} &= \sum_{\lm}
    \left\{
    X^{(+)}_{\lm} \left[ \frac{\pd^2}{\pd \theta^2} + \frac{1}{2} \ell(\ell + 1) \right] Y_{\lm}
    \right.
    \nonumber \\
    &\qquad \quad \left.
        - X^{(-)}_{\lm} \frac{\ii m}{\sin\theta} \left[ \frac{\pd}{\pd \theta} - \frac{\cos\theta}{\sin\theta} \right] Y_{\lm}
    \right\},
    \label{eq:h_plus} \\
    r \, h_{\times} &= \sum_{\lm}
    \left\{
    X^{(+)}_{\lm} \frac{\ii m}{\sin\theta} \left[ \frac{\pd}{\pd \theta} - \frac{\cos\theta}{\sin\theta} \right] Y_{\lm}
    \right.
    \nonumber \\
    &\qquad \quad \left.
        - X^{(-)}_{\lm} \left[ \frac{\pd^2}{\pd \theta^2} + \frac{1}{2} \ell(\ell + 1) \right] Y_{\lm}
    \right\},
    \label{eq:h_cross}
\end{align}
\end{subequations}
where $Y_{\ell m}$ are spherical harmonics, while $r$ corresponds to the distance to the black hole. In our case, $r = r_{\rm ext}$.
\endgroup

How does the EFT affect this result? Because the EFT corrections to the
propagation speed of the perturbations [cf.~Eq.~\eqref{eq:sounds_speed}] and
effective potentials [cf.~Eq.~\eqref{eq:effective_potentials_eft_schematic}]
are only significant in the near horizon region,
\emph{Equation~\eqref{eq:polarizations} holds in the EFT as well;
all imprints of the EFT corrections are encoded in the master functions
$X_{\lm}^{(\pm)}$ evaluated at future null infinity after they have scattered
off the EFT-corrected potential barrier.}
We verified that this is the case by inspecting the field equations
in the limit of large $r$, where one recovers the result of general
relativity~\cite{Martel:2005ir}.

\subsection{Synthetic signal generation: a toy model}
\label{subsec:toy_model}

Is there a way of using the results of Sec.~\ref{sec:simulations} to
obtain the gravitational-wave polarizations? In a self-consistent calculation,
each multipole $X^{(\pm)}_{\lm}$ would be computed by solving the inhomogeneous
version of Eq.~\eqref{eq:eqs_rwz} with the right-hand side of these equations being
given source terms that are responsible for driving the perturbations. For example,
these source terms could represent a particle plunging in geodesic motion
into the black hole, as studied in the classic papers by Davis et al.~\cite{Davis:1971pa,Davis:1971gg,Ruffini:1973ky}
in the 1970s, among others.
We leave this ongoing calculation in the EFT to a forthcoming paper in this
series~\cite{Tambalo:2026prep}.

Instead, for an initial study, we will make three assumptions that will enable us to
use the results from Sec.~\ref{sec:simulations}.
First, because of the spherical symmetry of our simulations in
Sec.~\ref{sec:simulations}, we will assume that our waveform $X_{\lm}^{(\pm)}$
was obtained as if we had evolved a momentarily static initial data with Gaussian
profile:
\begin{equation}
    \left. X_{\lm}^{(\pm)} \right\vert_{t = 0}
    =
    A_{\lm} \, \ee^{-(\rt - r_{\ast}^{\rm med})^2 / (2 \sigma^2)} \, \cos(m\phi),
\end{equation}
where $A_{\lm}$ is the amplitude. Spherical symmetry guarantees that the
dependence on the azimuthal angle $\phi$ will not affect the resulting
waveform~\cite{Glampedakis:2001js}.
Second, because $h_{+,\times}$ is real, we must include the terms $\pm m$ ($|m|
\leq \ell)$ when carrying out the summation in Eq.~\eqref{eq:polarizations}.
We will consider $\ell = 2$ only and assume that the amplitudes $A_{2m}$ are as follows:
\begin{equation}
    A_{22} = A_{2,-2} = 1, \quad \textrm{and} \quad A_{21} = A_{2,-1} = A_{20} = 0.
\end{equation}
Together with spherical symmetry, they imply that
\begin{equation}
    X_{22}^{(\pm)} = X_{2,-2}^{(\pm)} \equiv X_{2}^{(\pm)}
\end{equation}
are the only nonzero contributions to Eq.~\eqref{eq:polarizations}.
Having made this many assumptions, we do not shy away from making a
third one: to simplify the forthcoming analysis we will study only one polarization at a
time, which we choose to be $h_+$ for the most part of our discussion.

Under these assumptions, Eq.~\eqref{eq:h_plus} reduces to:
\begin{align} \label{eq:h_plus_final}
r \, h_+
&= f^{(+)}(\theta,\phi) \, X_{2}^{(+)}
\,+\,
f^{(-)}(\theta,\phi) \, X_{2}^{(-)},
\end{align}
where we defined
\begin{subequations} \label{eq:ang_form_factors}
\begin{align}
f^{(+)} &= \tfrac{1}{4} \sqrt{\tfrac{15}{2 \pi }} \, [3 + \cos 2 \theta] \cos 2 \phi,
\\
f^{(-)} &= \sqrt{\tfrac{30}{\pi }} \, \cos\theta \, \sin \phi \, \cos \phi,
\end{align}
\end{subequations}
and $r$ is a location far from the black hole that we take to be the
extraction radius $r_{\rm ext}$.
We notice that once we fix the angles $\theta$ and $\phi$, the ``form factors''
$f^{(\pm)}$ are degenerate with the amplitude $A_{\lm}$ of our initial data~\eqref{eq:id}.
Moreover, $h_+$ is completely determined by the parity-even contribution if
$\theta = \pi/2$ or when $\phi = 0$, $\pi/2$, $\pi$, $3\pi/2$, or $2\pi$.
Conversely, $h_+$ is completely determined by the parity-odd contribution if $\phi = \pi / 4$
or $3\pi/4$.

To analyze how isospectrality breaking affects $h_+$, we choose $\varepsilon =
0.05$ and study three pairs of angles $(\theta, \phi)$ as summarized next:
\begin{enumerate}[wide, labelwidth=!, labelindent=0pt]
    \item \emph{Comparable mixing}: we set $(\theta, \phi) = (\pi / 3,\, \pi/6)$.
        For these angles, $f^{(+)} \approx f^{(-)} \approx 0.67$.
        Thus, axial and polar waveforms $X^{(\pm)}$ have comparable contributions to $h_+$.
        \label{itm:comp_case}
    \item \emph{Polar dominated}: we set $(\theta, \phi) = (2 \pi / 5,\, \pi/10)$.
        For these angles, $f^{(+)} \approx 1.03$ and $f^{(-)} \approx 0.28$.
        Thus, the polar waveform $X^{(+)}$ is the dominant contribution to $h_+$.
    \item \emph{Axial dominated}: we set $(\theta, \phi) = (\pi,\, \pi/3)$.
        For these angles, $f^{(+)} \approx -0.39$ and $f^{(-)} \approx -1.34$.
        Thus, the axial waveform $X^{(-)}$ is the dominant contribution to $h_+$.
\end{enumerate}

\begin{figure*}[t]
    \includegraphics[width=\columnwidth]{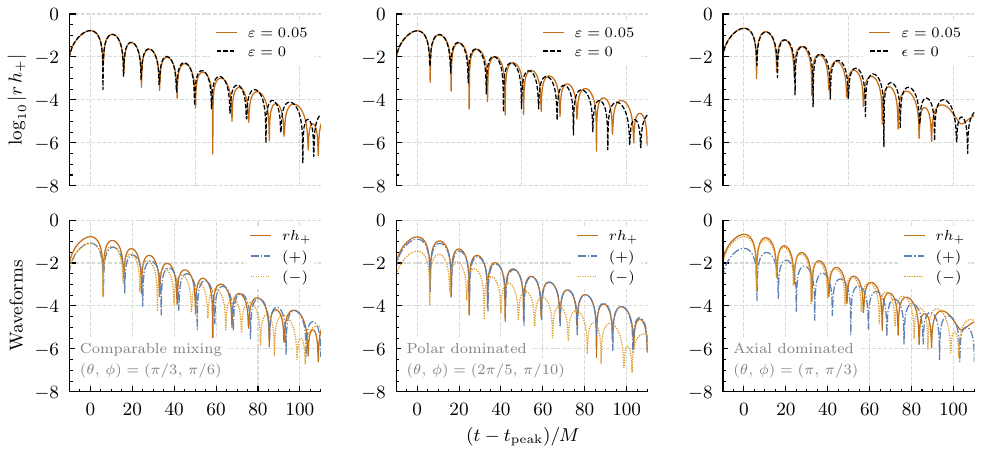}
    \caption{
        Gravitational waveforms. From left to right, we show the
        comparable-mixing, polar-dominated, and axial-dominated cases,
        respectively. In the top row, we compare $r h_+$ in the EFT with
        $\varepsilon = 0.05$ (solid line) and general relativity $\varepsilon =
        0$ (dashed line). In the bottom row, we compare $r h_+$ (for
        $\varepsilon = 0.05$, solid line) versus the polar [$(+)$, dot-dashed
        line] and axial [$(-)$, dashed line] master-function waveforms. We shifted all
        waveform in time such that their peaks occur at $t = \tp$.
    }
    \label{fig:all}
\end{figure*}

In Fig.~\ref{fig:all}, we show the resulting waveforms for each of these
cases. All waveforms are shifted in time such that their peaks occur at $t =
\tp$.
Moving left to right, the columns show the comparable-mixing, polar-dominated,
and axial-dominated cases. In the top row, we compare the gravitational-wave
polarization $h_+$ in the EFT with $\varepsilon = 0.05$ (solid line) against
general relativity $\varepsilon = 0$ (dashed line).
For each column, we show in the bottom row $h_+$ (again, for $\varepsilon =
0.05$, solid line) versus the polar [$(+)$, dot-dashed line], and axial [$(-)$,
dashed line] master-function waveforms. The latter two were multiplied by the
respective form-factors $f^{(\pm)}$.

A common trend can be seen in the panels in the bottom row. Let us start with the
comparable-mixing case shown in the leftmost bottom panel.
As expected, we see that at the start of the ringdown both axial and polar
waveforms contribute equally to $h_+$. However, as time progresses, $h_+$
becomes more similar to the polar waveform $X^{(+)}$; compare $h_+$ and $X^{(+)}$
in the intervals $0 \lesssim t/M \lesssim 50$ and $50 \lesssim t/M \lesssim
100$ after the peak of the waveform $\tp$.
This happens because the axial fundamental quasinormal frequency
$\omega_{20}^{(-)}$ has a smaller imaginary part than its polar counterpart;
see Eq.~\eqref{eq:qnm_20_eft}. For this reason, $X^{(-)}$ decays faster in
time, eventually leaving $X^{(+)}$ as the dominant contribution to ringdown.
This effect is also visible in the polar-dominated case shown in the middle bottom panel.
We see that the small difference between $h_+$ and $X^{(+)}$ at $t = t_{\rm
peak}$ decreases as the ringdown progresses, and the two curves become
virtually identical approximately $50M$ after $\tp$.
The opposite behavior happens in the axial-dominated case shown in the
rightmost bottom panel. In this case, $h_+$ and $X^{(-)}$ are very similar to one another
at the beginning of the ringdown, and differences between the two become
visible at $50M$ after $\tp$.

What about the comparison against general relativity?
In the comparable-mixing case (leftmost top panel), $h_+$ is virtually
indistinguishable initially in the EFT and in general relativity. Differences between the
two waveforms only become visible at about $50M$ after $\tp$. We
interpret this result as follows. Early in the ringdown, the polar and axial
fundamental quasinormal frequencies combine into an ``effective frequency''
\begin{equation} \label{eq:qnm_eff}
    M\omega^{\rm eff}_{20} = \tfrac{M}{2} \left[ \omega_{20}^{(+)} + \omega_{20}^{(-)} \right]
    \approx 0.376639 - 0.0876343\ii,
\end{equation}
which is approximately equal to the Schwarzschild fundamental quasinormal frequency~\eqref{eq:qnm_20}.
The axial mode decays faster in time as discussed
previously, causing $\omega^{\rm eff}_{20} \approx \omega_{20}^{(+)}$ as time progresses.
Indeed, the differences between the two waveforms become apparent $50 M$
after $\tp$.
In the polar- and axial-dominated situations (middle and right panels in the top row,
respectively) the EFT corrections are, to some extent, visible earlier in the
ringdown, as expected.

\subsection{Quasinormal frequency extraction}

Having constructed the waveforms $h_+$, we now explore to what
extent we can extract the \emph{two fundamental quasinormal frequencies} that
are predicted by EFT.

\subsubsection{Theory-agnostic single-frequency fits}

We begin by considering the same theory-agnostic single-frequency fit we
applied to $X^{(\pm)}$ in Sec.~\ref{sec:simulations:eft:qnm_fit}, but now we
applied to $h_+$.
As a sanity check, we first used this fit to the comparable-mixing case, but in
general relativity. Unsurprisingly, we were able to recover the fundamental
Schwarzschild quasinormal frequency~\eqref{eq:qnm_20} to within 1\% for $t_0 \geqslant 10 M$.
Our results are very similar to those shown in Figs.~\ref{fig:l2n0_fits}
and~\ref{fig:l2n0_fits_other_parameters}.

\begin{figure}[t]
  \includegraphics[width=1\columnwidth]{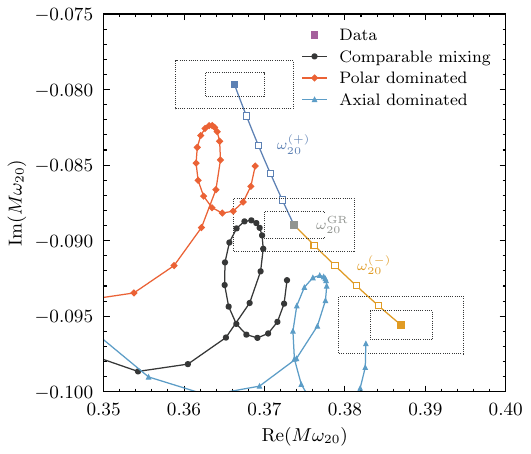}
  \caption{
      Trajectories in the complex plane for the best-fit fundamental
      quasinormal frequencies $\omega_{20}$ recovered from $h_+$, as $t_0$
      varies from zero to $30M$.
      As in Fig.~\ref{fig:l2n0_fits}, the squares indicate quasinormal frequencies
      in general relativity and in the effective field theory.
      The dotted-lined rectangles represent $\pm1\%$ and $\pm2\%$ away from
      reference value of $\omega_{20}$ when $\varepsilon = 0$ [Eq.~\eqref{eq:qnm_20}]
      and $\varepsilon = 0.05$ [Eq.~\eqref{eq:qnm_20_eft}.]
      The trajectories show that we are unable to accurately recover either of
      two fundamental quasinormal frequencies present in $h_+$, strikingly,
      even in the polar and axial dominated cases.
}
  \label{fig:fit_hp_single_mode}
\end{figure}

We then applied the same fitting procedure but now for $\varepsilon = 0.05$ and
the three cases enumerated in Sec.~\ref{subsec:toy_model}. The results
are summarized in Fig.~\ref{fig:fit_hp_single_mode}.
Similarly to Fig.~\ref{fig:l2n0_fits}, we show the trajectories in the complex
plane for the best fit quasinormal frequencies $\omega_{20}$ recovered from $h_+$
as we varied the start of the fit time-window $t_0$ from zero to $30M$.
We were unable to recover either of the fundamental
quasinormal frequencies~\eqref{eq:qnm_20_eft} in the three cases.
In the polar and axial-dominated cases, the curves whirl toward the region in
the complex plane where the polar and axial quasinormal frequencies move away
from their general-relativistic values as $\varepsilon$ increases. While the
two trajectories do seem to converge, they do so at none of the frequencies of the
black hole. Our best-fit frequencies are thus biased.
Because \emph{we were} able to recover the fundamental quasinormal frequencies
when we analyzed the individual $X^{(\pm)}$ functions, we conclude that the
\emph{mixing of frequencies in $h_+$ is the cause of this result.}
Interestingly, we see that the trajectory for the comparable-mixing case whirls
toward the Schwarzschild quasinormal frequency, being 2\% away from this
frequency for $13 \lesssim t_0/M \lesssim 19$. This observation supports our
interpretation from Sec.~\ref{subsec:toy_model} that at least during some time
in the ringdown the two EFT
fundamental quasinormal frequencies behave as a single effective
frequency~\eqref{eq:qnm_eff} that, in the EFT, has a numerical value close to
the general-relativity value~\eqref{eq:qnm_20}.

The deterioration of our ability to recover the frequencies in the polar and
axial dominated cases is qualitative similar to the conclusions of
Ref.~\cite{Volkel:2025jdx}.

We also experimented using a two-frequency fit to analyze $h_{+}$, with the
hope that we could accurately extract either of the two fundamental quasinormal
frequencies from the waveform. We used Eq.~\eqref{eq:qnm_fit} with $N=1$,
increasing the number of free parameters in our model from four to eight.
We attempted to fit all parameters simultaneously. The result was disastrous: neither
frequencies converged as we varied the time-window parameter $t_0$.

In light of these sobering results, we turn to a (perhaps) more modest
question: can we infer the presence of a nonzero value of $\varepsilon$?

\subsubsection{Can we infer a nonzero $\varepsilon$?}

After having performed theory-agnostic fits for the waveform $h_{+}$, we now discuss an EFT-informed fit.
This means that the complex frequency of the fundamental mode is not left free to vary but is assumed to take the form predicted by our EFT: $\omega_{\ell n}^{(\pm)} = \omega_{\ell n}^{\rm GR} +\varepsilon \, \delta \omega_{\ell n}^{(\pm)}$. For the case of $\ell = 2$ and $n = 0$, the linear correction to the frequencies can be evaluated to be
\begin{subequations} \label{eq:delta_omega_EFT}
\begin{align}
  M \delta \omega_{2 0}^{(+)} & = -0.145410 + 0.166074 \ii,
  \label{eq:delta_omega_EFT_even}
  \\
  M \delta \omega_{2 0}^{(-)} & = + 0.250595 - 0.133418 \ii .
  \label{eq:delta_omega_EFT_odd}
\end{align}
\end{subequations}
In order to avoid confusion, we refer to $\omega^{(+)}$ in Eq.~\eqref{eq:delta_omega_EFT_even} as the \emph{even} mode and to $\omega^{(-)}$ in Eq.~\eqref{eq:delta_omega_EFT_odd} as the \emph{odd} mode. These frequencies correspond, respectively, to the polar and axial quasinormal-frequencies.
Notice that the even mode is longer-lived compared to the odd mode.

With this procedure, we can test whether the two parities contained in the waveform can be extracted,
so to improve the two-frequencies fit.
As we will see, even with this procedure, the extraction of the two modes remains difficult.
On the other hand, we will show that, for configurations dominated by a single parity mode, our fits give a preference for the correct parity.
However, for EFT waveforms, the best-fit value for the EFT parameter $\varepsilon$ does not
always align with the correct value.
Below we will discuss the results of our fits, obtained by following the same procedure as in Sec.~\ref{sec:simulations:eft:qnm_fit}.
In the least-square fitting, we utilize the same bounds for the amplitude and phase, while imposing the
following bound on the EFT parameter $\varepsilon \in [0, 0.2]$.\footnote{To avoid too stringent priors on
$\varepsilon$, we chose the upper limit to be $0.2$, which is outside of the regime where EFT corrections are linear \cite{Silva:2024ffz}.
However, we checked that changing this value does not alter our results.}

\begin{figure}[t]
\includegraphics{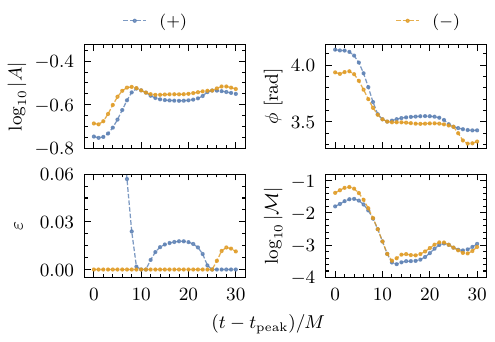}
\caption{Dependence of the mismatch and the best-fit parameters in our fitting model as functions of $t_0 = t - \tp$ for the waveform $h_+$ in general relativity obtained using Eq.~\eqref{eq:h_plus_final} in the comparable-mixing configuration and the master function shown in Fig.~\ref{fig:axipolar}, assuming an EFT form for the quasinormal-mode frequencies.
}
\label{fig:fit_eftinfo_gr}
\end{figure}

First, we consider a waveform in general relativity with comparable polar and axial components, i.e., the ``comparable mixing'' case of Sec.~\ref{subsec:toy_model}. We then perform a single-mode fit, first with a ``even-mode'' assumption for the frequency \eqref{eq:delta_omega_EFT_even} and then with an ``odd-mode'' assumption \eqref{eq:delta_omega_EFT_odd}.
We show the results for the two fits in Fig.~\ref{fig:fit_eftinfo_gr}. Both models (even and odd parametrizations of the fundamental frequency) produce similar results for the amplitude $|A|$, phase $\phi$ and mismatch $\mathcal M$. Regarding $\varepsilon$, we notice that the odd-mode fit consistently obtains $\varepsilon \simeq 0$, while the even-mode fit prefers a nonvanishing (but small) value for this parameter.
Among the two fits, the best overall fit is achieved for the even model when
$(t - \tp) \simeq 12M$ and with $\varepsilon \simeq 0.01$. This result
establishes the sensitivity of our analysis, indicating that a value of
$\varepsilon \simeq 0.01$ is the smallest deviation from general relativity our model
can confidently distinguish from zero.

As a second test, we consider an EFT waveform with $\varepsilon = 0.05$ in two configurations: the \emph{polar} and \emph{axial dominated} cases of \ref{subsec:toy_model}. In the two angular configurations, we fit the waveform $h_+$ with both even \eqref{eq:delta_omega_EFT_even} and odd-mode \eqref{eq:delta_omega_EFT_odd} models.
For simplicity, for each angular configuration, we keep the best fit between even- and odd-mode models.
The goal in these cases is to find a preference for the correct parity and recover the correct value for $\varepsilon$.

\begin{figure}[t]
\includegraphics{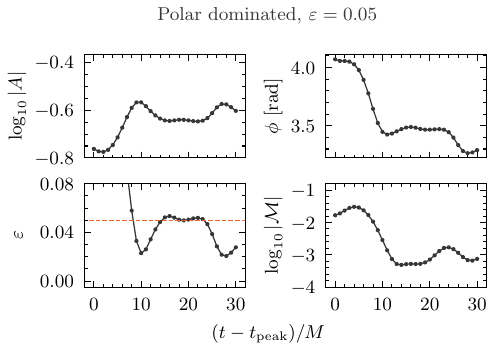}
\includegraphics{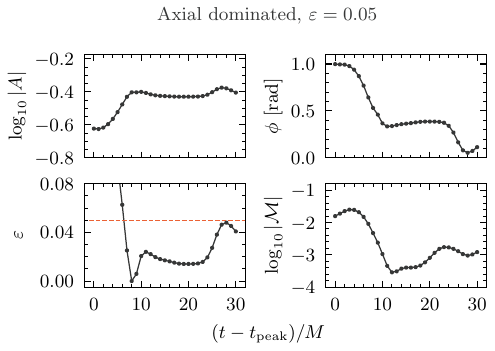}
\caption{
Dependence of the mismatch and the best-fit parameters in our fitting model as
functions of $t_0 = t - \tp$ for the EFT waveform $h_+$ obtained using
Eq.~\eqref{eq:h_plus_final} in the polar-dominated (top set of panels) and
axial-dominated (bottom set of panels) configurations.
We set $\varepsilon = 0.05$ in both cases; this value is indicated by the
horizontal dashed line in the lower-left panels.
In each case, the smallest mismatches are obtained when we use the even (odd)
ansatz for the quasinormal frequency~\eqref{eq:delta_omega_EFT} to analyze
the polar dominated (axial dominated) configuration, as expected.
}
\label{fig:fit_eftinfo}
\end{figure}

We show our results for the polar and axial-dominated configurations in the top
and bottom set of panels in Fig.~\ref{fig:fit_eftinfo}, as indicated therein.
For the polar-dominated configuration, we obtain the smallest mismatch when starting
the fit around $(t - \tp) \simeq 12M$ after the peak. The best fitting model is, as
expected, the even-mode model. Moreover, the recovered value for $\varepsilon$
(best fit) is approximately $0.049$, very close to the true value, $0.05$,
indicated by the dashed horizontal line.
For the axial-dominated configuration, the smallest mismatch is again obtained starting
approximately at $12M$ after the peak, with the odd-mode model correctly
identified. Contrary to the polar-dominated configuration, the
recovered value for $\varepsilon$, around $0.022$, is not as precise.

Let us comment on the results of the fits in these two angular configurations.
First, the fitting procedure provides stable results. Indeed, in all cases, within the time window between $10$ to $20 M$ after the peak, the fits return constant values for the parameters. In this region, the mismatch is also minimized.
Second, the polar-dominated configuration provides a better reconstruction for $\varepsilon$. This result can be understood by recalling that, in the EFT, the polar mode is longer-lived than the axial mode. This means that it is easier to fit for the polar mode, as it contributes more at later times. On the other hand, the axial mode decreases more quickly and is thus less ``visible.''

In addition to the two fits above, we have also tested the EFT-informed fit in the ``comparable mixing'' configuration, where both parities are similarly excited, with $\varepsilon = 0.05$. Here, the fits slightly prefer the even-mode model, with a poorly reconstructed value of $\varepsilon \simeq 0.01$. Given the general poor reconstruction for $\varepsilon$, we did not attempt to perform an EFT-informed two-mode fit, i.e.,~fitting simultaneously for two amplitudes and phases and $\varepsilon$, assuming the two frequencies follow Eqs.~\eqref{eq:delta_omega_EFT_even} and \eqref{eq:delta_omega_EFT_odd}.

As an additional test for our fits, we have repeated them with $h_\times$ instead of $h_+$. The former waveform is obtained using Eq.~\eqref{eq:h_cross} (also in this case, we consider $\ell = 2$ and the sum over $m = \pm 2$).
From this formula, we notice that, compared to $h_+$, the roles of $X^{(+)}_{\ell m}$ and $X^{(-)}_{\ell m}$ are switched, and the overall sign is flipped.
This means in particular that a polar-dominated angular configuration for $h_+$ corresponds to an axial-dominated configuration for $h_\times$.
We then find consistent results with the previous analysis: in polar-dominated configurations for $h_\times$, we are able to recover the parameter $\varepsilon$ accurately and identify the correct mode, while in axial-dominated configurations, $\varepsilon$ is poorly reconstructed, but the mode is identified correctly.

Finally, we end by commenting on the case where $\varepsilon$ is not imposed to be positive, but allowed to take any sign.
In this case, we find that the fits generically degrade, and in particular, polar (axial) dominated configurations can be fitted better by the odd (even) model.
This can be understood by noticing the presence of an approximate degeneracy when sending $\varepsilon \to - \varepsilon$ and by exchanging the mode parity. Indeed, by inspecting Fig.~\ref{fig:l2n0_fits}, one sees that by continuing the straight trajectory for, say, $\omega^{(+)}$ to negative $\varepsilon$, we remain close to the $\omega^{(-)}$ trajectory with $\varepsilon > 0$. Therefore, fitting the waveform with $\omega^{(+)}$ and $\varepsilon < 0$ yields similar mismatches as with $\omega^{(-)}$ and $\varepsilon >0$. This further highlights the difficulty in correctly identifying deviations from general relativity, and in particular, isospectrality breaking, in realistic ringdown data.

\section{Conclusions and outlook} \label{sec:conclusions}

We presented a first study of  how isospectrality breaking affects the
time-domain gravitational-wave polarizations in the ringdown of perturbed
nonrotating black holes in an EFT extension of general relativity.
We began by evolving in time the master equations that dictate the dynamics of
polar and axial perturbations using momentarily static Gaussian initial data.
We then proposed a prescription on how to combine the resulting polar and
axial-perturbation waveforms to construct the complete gravitational waveform.
With these waveforms in hand, we explored to what extent we can distinguish the
presence of the two fundamental quasinormal frequencies in the signal, the chief
implication of the loss of isospectrality.

Our main conclusions are clear. First, it is hard to infer the presence
of these two modes when using a model agnostic fit for the ringdown. The
presence of a doublet of fundamental quasinormal frequencies in the signal
deteriorates our ability to recover either.
Second, using a ``EFT-informed'' model (in which the damping and oscillation
frequency of the quasinormal modes are written as a perturbative
expansion in the EFT parameter $\varepsilon$ around their Schwarzschild values), however,
we were at least able to infer the presence of a deviation of general
relativity in the signal. Still, this is only possible when the ringdown is
dominated by the polar fundamental mode, which has the slowest damping time from the doublet.

We entertain the possibility that similar conclusions will be shared in the ringdown
of nonrotating black holes in other theories that are perturbatively close to general
relativity and possess only massless degrees of freedom. Examples include EFTs of general relativity
with quartic-in-curvature terms and theories that involve nonminimal coupling of (pseudo)scalar
fields with curvature scalars, such as shift-symmetric Einstein-scalar-Gauss-Bonnet and dynamical
Chern-Simons gravity.

How do our conclusions depend on our assumptions?
First, while the relation~\eqref{eq:polarizations} between $h_{+,\times}$ and
the ZM (polar) and CPM (axial) master functions hold in the EFT, there is
clearly room to improve how the latter were computed.
As discussed in Sec.~\ref{subsec:toy_model}, a self-consistent calculation of
the gravitational waveforms can be made using a test particle, whose geodesic
motion excites the black-hole perturbations. This problem will be tackled
in an upcoming paper in this series~\cite{Tambalo:2026prep}.
Still, it is tempting to conjecture that our main conclusions will be shared in such a
calculation. This is justified by the flexibility of our construction leading to
Eq.~\eqref{eq:h_plus_final} in accommodating the relative contributions of the
axial and polar contributions to the waveform.
Second, we have considered only nonrotating black holes. Quasinormal frequency calculations
of rotating black holes in the EFT considered herein show that the fundamental polar and axial
quasinormal frequencies can deviate more from their general-relativity values for larger black hole
spins~\cite{Cano:2023jbk,Cano:2024jkd}. It would be important to understand how rotation affects our findings.

In summary, there is much to be done yet to understand the quasinormal modes, their excitation,
and observational implications in theories beyond general relativity.

\section*{Acknowledgements}
We thank Barry Wardell for the suggestion of carrying out scattering experiments to
study the loss of quasinormal mode isospectrality in our problem and Helvi
Witek for numerous discussions and advice.
We thank Mark Ho-Yeuk Cheung for correspondence and sharing with us data from
Ref.~\cite{Baibhav:2023clw} which we used to validate our fitting routines.
We also acknowledge discussions with Jamie Bamber, Caio F.~B. Macedo, Jan
Steinhoff, Antonios Tsokaros, Sebastian V\"olkel, and Daiki Watarai.
Some of our calculations were done with the {\sc Mathematica} packages {\sc
xPert}~\cite{Brizuela:2008ra} and {\sc
Invar}~\cite{Martin-Garcia:2007bqa,Martin-Garcia:2008yei}, parts of the {\sc
xAct/xTensor} suite~\cite{Mart_n_Garc_a_2008,xAct}.
H.O.S. acknowledges funding from the Deutsche Forschungsgemeinschaft
(DFG) - Project No.:~386119226.
K.Y. acknowledges support from NSF Grant PHY-2309066 and PHYS-2339969.

\appendix

\section{Code details and validation} \label{app:code_description}

In this appendix, we provide details of our numerical code and
describe the convergence tests we performed to validate it.
We evolved Eq.~\eqref{eq:first_order_pdes} using the method of lines.
In particular, our code uses fourth order accurate-in-space finite difference
stencils and evolve the variables $X$, $\Psi$, and $\Pi$ forward in time using
the third-order accurate Rugge--Kutta scheme of Shu and
Osher~\cite{Shu:1988JCoPh..77..439S}.
We also implemented, and tested, a fourth-order accurate Rugge--Kutta scheme.
We opted to use the third-order scheme as a compromise between accuracy
and simulation run times.
The code is written in C++ and uses OpenMP for parallelization.

\subsection{Grid size and boundary conditions}

In all our simulations we used a uniform spatial grid ranging from
$r_{\ast}^{\rm min} = -30M$ to $r_{\ast}^{\rm max} = 600M$, while time
integration was carried out from $t = 0$ to $t = 720M$.
Our choices for the grid parameters require some justification.
Let us begin with the spatial domain. Because we employ tortoise coordinates in
our spatial grid, we must compute numerically the inverse relation
\begin{equation} \label{eq:r_fun_rt}
r=r(\rt; \, \varepsilon),
\end{equation}
to evaluate the propagation speed and effective potential in
Eq.~\eqref{eq:eqs_rwz}. The explicit expression for $\rt = \rt(r;\, \varepsilon)$
can be found in Ref.~\cite{Silva:2024ffz}, Appendix D.

\begin{figure}[t]
\includegraphics{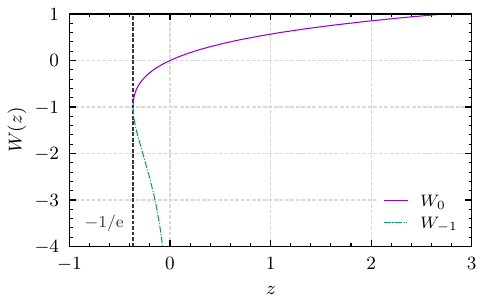}
\caption{The two real branches of the Lambert $W$ functions, $W_{0}$ (solid line) and $W_{-1}$ (dot-dashed line). The principal branch $W_0$ is defined by $-1 \leq W$, this occurs for $-1/{\rm e} \leq z$ (dashed gray line).}
\label{fig:lambert}
\end{figure}

In general relativity, $\varepsilon = 0$, the two coordinates are related as
follows:
\begin{equation} \label{eq:rt_lambert}
    \rt = r + 2 M \ln|r/(2M) - 1|.
\end{equation}
Equation~\eqref{eq:rt_lambert} can be inverted using the Lambert $W$
function~\cite{Corless:1996zz,Boonserm:2008zg}.
This function is defined as the solution to the equation
\begin{equation} \label{eq:lambert}
    W(z) \, \ee^{W(z)} = z.
\end{equation}
In our problem, we find that
\begin{equation} \label{eq:r_fun_rt_gr}
r = 2M \, [ 1 + W(z) ], \qquad z = \exp[\rt/(2M) - 1].
\end{equation}
The function $W(z)$ has two real values in the interval $-1/\ee \leq z < 0$;
see Fig.~\ref{fig:lambert} for an illustration.
In particular, at $z=-1/\ee$, $W(z) = -1$.
The principal branch of the Lambert $W$ function is defined by $-1 \leq
W(z)$. In our case $z \in [0, \infty)$, and this is the branch we must use.

When $\varepsilon \neq 0$, we calculate~\eqref{eq:r_fun_rt} numerically by
locating the root of $\varrho(\rt;\, \varepsilon) = r - r(\rt; \varepsilon)$
To do this, we first compute a value of $r$ using the general relativistic
result~\eqref{eq:r_fun_rt_gr}. Call it $r_0 = r(\rt;\,0)$.
We then calculate $r_{\rm L} = r_0 (1 - \delta r)$ and
$r_{\rm R} = r_0 (1 + \delta r)$ that serve as educated guesses to bracket
the root, which we determine through bisection.
We found that $\delta r = 0.1$ works for this task for all values of
$\varepsilon$ considered in this paper.
When computing $r_{\rm L}$ we must ensure that it is not located inside the
event horizon, $\rh$. To avoid this edge case, we first check if $r_{\rm L}
\geq \rh$. If this is true, we use the foregoing expression for $r_{\rm
L}$. If false, we set $r_{\rm L} = \rh (1 + 10^{-10})$.
In practice, our code uses the implementations for the Lambert $W$ function and
bisection algorithm from the Boost C++ libraries~\cite{Boost}.

When setting up the numerical grid in $\rt$, we must be careful with the fact
that $r/\rh$ is indistinguishable from $1$ at double precision for any
$\rt / \rh \lesssim -30$ when $\varepsilon = 0$, for which $\rh = 2M$.
In other words, the tortoise coordinate $r(\rt)/\rh$ will evaluate exactly
to $1$ for any value of $\rt/\rh$ smaller than approximately $-30$.
For this reason, we restrict ourselves
to $\rt^{\rm min}/\rh > - 30$. In general relativity, this translates into
$r(\rt^{\rm min})/\rh - 1 \gtrsim 3.5 \cdot 10^{-14}$,
which gives us a rule of thumb of how slightly outside the event horizon our numerical grid can begin.
In practice, we choose $\rt^{\rm min} = - 30M$. The reason is twofold. First,
it satisfies the above bound. Second, for this value, both $c^{2}_{\rm s}$ and
$V^{(\pm)}_{\ell}$ are sufficiently close to zero for us to impose outgoing
boundary conditions at the edge of our numerical grid as if we had a simple
one-dimensional wave equation propagating with constant velocity.
This partly justifies our choice of $\rt^{\rm max} = 600M$ too. The remaining
justification, and the explanation of our choice of time integration range, is
that we wish to avoid numerical reflections from the boundaries to appear in
our waveforms given the parameters of the initial data. We give more details below.

\subsection{Initial data and waveform extraction}

We set the parameters of the Gaussian initial data~\eqref{eq:id} to be
\begin{equation}
    A = 1,
    \quad \sqrt{2} \, \sigma = 1.5M,
    \quad \textrm{and} \quad r_{\ast}^{\rm med} = 100M.
\end{equation}
The width $\sigma$ of the Gaussian is chosen to be smaller than the
lengthscales associated with the spatial extension in which the
propagation speed $c^{2}_{\rm s}$ and the effective potential $V^{(\pm)}_{\ell}$
are nonzero. In this way, we can observe the effects of $c^{2}_{\rm s}$ and
$V^{(\pm)}_{\ell}$ in the scattered wave.

The value of the scattered wave, that is, the waveform, is extracted at a radius
$\rt^{\rm ext} \geq \sigma$ of ${\cal O}(10^{2}M)$
For these values of $\rt^{\rm ext}$, $\sigma$, and $\rt^{\rm min / max}$, we
avoid the contamination in our waveforms from numerical noise of small
amplitude originating from reflected waves at the boundaries of the spatial
grid in the duration of our simulations.

\subsection{Convergence tests and error estimates}

\begin{figure}[t]
\includegraphics{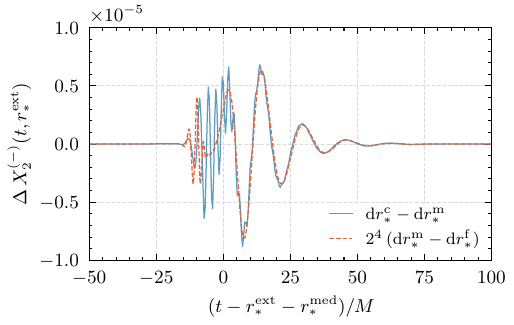}
\caption{Local convergence analysis of the quadrupolar CPM waveform extracted at $r_{\ast}^{\rm ext} = 150M$,
computed with different spatial resolutions $\dd r_{\ast}$ and fixed Courant factor $C = 0.2$. The figure shows
the difference between the waveforms obtained using coarse ($\dd r_{\ast}^{\rm c} = 0.2M$) and medium ($\dd r_{\ast}^{\rm m} = 0.1M$)
resolutions (solid line) and medium and fine ($\dd r_{\ast}^{\rm f} = 0.05M$) resolutions (dashed line).
We shifted the curves in time by the extraction radius $r_{\ast}^{\rm ext}$
and initial location of the Gaussian's peak, $r_\ast^{\rm med} = 100M$.
We rescaled the latter curve by the expected convergence factor, $2^4$. The
overlap between the two curves demonstrates the fourth order convergence of the
code.}
\label{fig:convergence}
\end{figure}

In Fig.~\ref{fig:convergence}, we show the results of our local convergence test.
We chose a Courant factor $C = \dd t / \dd r_{\ast} = 0.2$.\footnote{Because the propagation speed $\cs$ has
spatial dependence, using the same Courant factor $C$ across the numerical domain may not be appropriate.
That is, we should be careful and include some information about
$\cs$, for instance, its maximum value, when choosing the Courant factor.
However, this is not an issue in our problem because we choose a small value of $C$, and $c^{2}_{\rm s}$ deviates from
unity by at most by 3\%; see Fig.~\ref{fig:speed}.}
We then performed three simulations with increasing spatial resolution, $\dd r_{\ast} = \{0.05,\, 0.1,\, 0.2\}$,
with fixed values of the effective-field-theory parameter $\varepsilon = 5 \times 10^{-2}$, considering the
Regge--Wheeler potential $V^{(-)}_{\ell}$, and quadrupolar $\ell = 2$ perturbations.
The waveform was extracted at $r_{\ast}^{\rm ext} = 100M$.
The solid and dashed curves shows the difference between the waveforms produced
in the coarse ($\dd r_{\ast}^{\rm f} = 0.2M$) and medium ($\dd r_{\ast}^{\rm f}
= 0.1M$) and medium and fine ($\dd r_{\ast}^{\rm f} = 0.05M$)
resolutions, respectively. We rescaled the latter curve by $2^4$, the expected
convergence factor in our test. The resulting overlap between the two curves
demonstrates the fourth-order convergence of the code.
All the simulations presented in the main text use the medium spatial
resolution and Courant factor $C = 0.2$.

To tell apart the effects of the effective-field-theory parameter
$\varepsilon$ from numerical noise, it is important to quantify the error
associated with our waveforms.
Because our code is in the convergence regime (as shown in
Fig.~\ref{fig:convergence}), we calculated the absolute value of the difference
between our coarse- and medium-resolution waveforms and divided by $2^4 -
1$~\cite{Alcubierre:2008itnr}.
Taking the absolute value and then the mean value of this quantity, we arrive
at the error estimate of ${\cal O}(10^{-6})$ in our waveforms. This is smaller
than the values of $\varepsilon$ we consider that are ${\cal O}(10^{-2})$.
Consequently, we can interpret the changes in the waveforms produced by varying
values of $\varepsilon$, while keeping the other simulation parameters fixed, as
originating from the higher-curvature corrections instead of numerical errors.

\begin{figure}[t]
\includegraphics{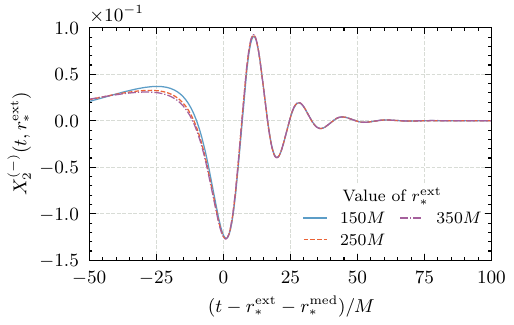}
\caption{Quadrupolar CPM waveform extracted at three different
radii $\rt^{\rm ext} = 150M$ (solid line), $250M$ (dashed line), and $350M$ (dot-dashed line).
We use a spatial grid resolution of $\dd \rt = 0.1M$
and a Courant factor $C=0.2$, i.e., the medium resolution setup in Fig.~\ref{fig:convergence}.
The waveforms are shifted in time according to where they were extracted,
$\rt^{\rm ext}$ and initial location of the Gaussian's peak, $r_\ast^{\rm med}
= 100M$.
The overlap between the waveforms demonstrates that numerical dispersion and dissipation are
small in the code.
}
\label{fig:convergence_different_radii}
\end{figure}

We also performed another check of our code.
Figure~\ref{fig:convergence_different_radii} shows the quadrupolar
Regge--Wheeler waveform from the medium resolution simulation of our local
convergence test, with the waveform extracted at three different radii, $\rt^{\rm
ext} = 150M$ (solid line), $250M$ (dashed line), and $350M$ (dot-dashed line).
Each waveform is shifted in time according to its extraction radius, $\rt^{\rm
ext}$, and initial location of the Gaussian's peak, $r_\ast^{\rm med}$.
We see that the three waveforms overlap, notably in the ringdown.
This demonstrates that numerical dissipation and dispersion are small in the code.
However, there is a small difference between the waveforms in the
(approximate) interval $(t - \rt^{\rm ext} - \rt^{\rm med}) / M \in [-50,\,0]$.
We attribute this difference to the proximity of the smallest extraction radius
to the initial data.
Indeed, we see that the difference between the waveforms decreases as $\rt^{\rm
ext}$ increases.

\section{The wave equation with a position-dependent propagation speed}
\label{app:schwarz}

\subsection{Change of variable to $c_s = 1$}
The evolution equation with a position-dependent speed of sound $\cs(x)$ can be
mapped into an equation with $\cs = 1$ and a modified potential; see
Eq.~\eqref{eq:eqs_rwz_schwarz} in the main text.
In this appendix, we follow the presentation in Ref.~\cite{Trachanas} and re-derive the transformed equation.
We start with the following frequency-domain equation
\begin{equation}\label{eq:psi_cs}
    \psi''(x) + \frac{\omega^2}{\css(x)} \psi(x) - V(x) \psi(x) = 0.
\end{equation}
To achieve our goal, we change variable from $x$ to $y$ and introduce a new field:
\begin{equation}\label{eq:def_phi_y}
\varphi(y) = \psi(x(y)) / \sqrt{\dot x(y)},
\end{equation}
where the overdot stands for the derivative with respect to $y$ (not to be confused with a time derivative).
Let us call $g(y) = \sqrt{\dot x}$.
Then, we find that
\begingroup
\allowdisplaybreaks
\begin{subequations}
\begin{align}
    g' &= \frac{1}{2g}\frac{\ddot x}{\dot x},
    \\
    g''&= \frac{1}{g^3}
        \left[
            \frac{1}{2} \frac{\dd}{\dd y}\left(\frac{\ddot x}{\dot x}\right)
            -
            \frac{1}{4}
            \left(\frac{\ddot x}{\dot x}\right)^2
        \right]
        \equiv \frac{1}{g^3} \{x, y\},
    \label{eq:gpp}
    \\
    \varphi' & = \frac{\dot \varphi}{g^2},
    \\
    \varphi''
    & =
    \frac{1}{g^4}
    \left[
        \ddot \varphi
        - \frac{\ddot x}{\dot x}
        \dot \varphi
    \right].
\end{align}
\end{subequations}
\endgroup
In Eq.~\eqref{eq:gpp} we identified the Schwarzian derivative:
\begin{equation}
    \{x, y\} =
    \frac{1}{2} \frac{\dd}{\dd y}\left(\frac{\ddot x}{\dot x}\right)
    -
    \frac{1}{4} \left(\frac{\ddot x}{\dot x}\right)^2 \,.
\end{equation}
Using these expressions in Eq.~\eqref{eq:psi_cs} we obtain
\begin{equation}
    \ddot \varphi
    +
    \left[
        g^4 \left(\frac{\omega^2}{\css(x(y))} -V(x(y))\right) + \{x, y\}
    \right]
    \varphi = 0 .
\end{equation}
Notice that by choosing $g^4(y) = \dot x^2 = \css(x(y))$, we are effectively
setting $\cs = 1$ for the field $\varphi$. This choice implies that
\begin{equation}
y(x) = \int^x {\rm d} x' / \cs(x').
\end{equation}
With this choice, we finally obtain
\begin{equation}\label{eq:phi_W}
    \ddot \varphi + [\omega^2 - W(y)] \, \varphi = 0,
\end{equation}
with the new potential
\begin{equation}
    W(y) = \css(x(y)) \, V(x(y)) - \{x, y\}.
\end{equation}

The Schwarzian derivative $\{x, y\}$ in terms of the function $\cs(x)$ takes the form
\begin{equation}
    \{x, y\} = \tfrac{1}{2} \cs \, \cs'' - \tfrac{1}{4}(\cs')^2.
\end{equation}
Equation~\eqref{eq:phi_W} can be Fourier-transformed back to the time domain,
resulting in a wave equation with uniform propagation speed
and potential $W$.
In the next section, we apply this procedure to the wave equation that occurs
in our problem, Eq.~\eqref{eq:eqs_rwz}.

\subsection{The metric-perturbation case}

The procedure outlined above can be applied to the case of the metric perturbation equations~\eqref{eq:eqs_rwz}. These equations indeed match Eq.~\eqref{eq:psi_cs} when identifying $x$ with the tortoise coordinate $\rt$.
Also, the metric perturbations $X^{(\pm)}_{\ell}$ correspond to $\psi$ and, by analogy, we define $Y^{(\pm)}_{\ell}$ as the field variables after the transformation.
Then, after the transformation, both axial and polar potentials $V^{(\pm)}_{\ell}$ get mapped into new potentials
\begin{equation}\label{eq:W_pot}
W^{(\pm)}_{\ell} = c^2_{\rm s}(\rt(\tilde r)) \, V^{(\pm)}_{\ell} - \{\rt, \rtl\},
\end{equation}
where we identify $\rtl$ with $y$. Note that the Schwarzian and the propagation-speed rescaling affect the two original potentials in the same way, with no $\ell$ dependence.\footnote{Let us assume for the moment the Schwarzian is the only correction to the potentials of general relativity.
Even though in general relativity $V^{(\pm)}_{\ell}$ are isospectral, a common shift by the function $\{\rt, \rtl\}$ does not necessarily lead to isospectral potentials.
}
When formulating the equations in terms of $Y^{(\pm)}_{\ell}$, the effect of the speed of sound $\cs$ can be quantified by evaluating the relative contribution of the Schwarzian to the potential $W^{(\pm)}_{\ell}$. In particular, we are interested in comparing this contribution with the EFT corrections already present in the potential $V^{(\pm)}_{\ell}$, given in Eq.~\eqref{eq:effective_potentials_eft_schematic}.

We start with the definition of the new coordinate $\rtl$, assigning $x = \rt$:
\begin{equation}
    \rtl(\rt) = \int^{\rt'} \, \dd \rt' \, \frac{1}{c_{\rm s}(\rt')}.
\end{equation}
For the purpose of the code implementation, it is useful to write $\rtl = \rtl(r)$.
We first perform a change of variables in the integral:
\begin{equation}
    \rtl(r) = \int^{r'} \, \dd r' \, \frac{\dd \rt'}{\dd r'} \frac{1}{c_{\rm s}(r')}
            = \int^{r'} \, \dd r' \, \frac{1}{a(r') \, c_{\rm s}(r')}.
\end{equation}
or in differential form:
\begin{equation} \label{eq:ode_dydr}
    {\dd \rtl}/{\dd r} = (a \, \cs)^{-1}, \quad a = N f.
\end{equation}
When $\cs = 1$, that is, when $\varepsilon = 0$, $\rtl$ is nothing but the tortoise coordinate.
Using the explicit forms of $a$ and $\cs$, we find
\begin{align}
    \frac{1}{a \, \cs} &= \left(1 - \frac{\rh}{r}\right)^{-1} \, \left\{ 1 +
    \varepsilon \left[ \frac{5}{8}\frac{M}{r} + \frac{5}{4}\frac{M^2}{r^2} + \frac{5}{2}\frac{M^3}{r^3} \right. \right.
    \nonumber \\
                             &\quad \left. \left. + \frac{5M^4}{r^4} + \frac{154M^5}{r^5}\left(1 - \frac{72}{77}\frac{\rh}{r}\right) - \frac{88M^6}{r^6} \right] \right\}.
    \nonumber \\
\end{align}
to leading order in $\varepsilon$. We integrate the differential
equation~\eqref{eq:ode_dydr} for $y$.
The solution has the schematic form:
\begin{equation}
    \rtl = r + \rh \log(r/\rh - 1) + \varepsilon \, \delta \rtl(r),
\end{equation}
where we chose the integration constant such that $\tilde{r}$ reduces to the
familiar tortoise coordinate in Schwarzschild when $\varepsilon = 0$. The
particular form of $\delta \rtl$ is not illuminating.

We can now compute the Schwarzian derivative $\{\rt, \rtl\}$:
\begin{align}\label{eq:schw_fun_r}
    \{\rt, \rtl \} &= \frac{1}{2} \cs \, \cs'' - \frac{1}{4} (\cs')^2,
    \nonumber \\
               &= \frac{1}{2} \cs \left( a^2 \, \frac{\dd^2 \cs}{\dd r^2} + a \frac{\dd a}{\dd r} \frac{\dd  \cs}{\dd r} \right)
               - \frac{1}{4} \left(a \frac{\dd \cs}{\dd r} \right)^2,
    \nonumber \\
               & = - \frac{2160M^5}{r^5} \frac{\varepsilon}{r^2}
               \Big(1 - \frac{\rh}{r}\Big)
               \Big(1 - \frac{16}{15}\frac{\rh}{r}\Big)
    \nonumber \\
               &\quad \times \Big(1 - \frac{3}{2}\frac{\rh}{r}\Big).
\end{align}
Note that $\{\rt, \rtl\}$ is not positive definite, has roots at
\begin{equation}
    r = \rh,\quad r = 16 \rh / 15, \quad \textrm{and} \quad r = 3 \rh / 2,
\end{equation}
and vanishes as $r$ approaches infinity.

To relate the variables $Y^{(\pm)}_{\ell}$ and $X^{(\pm)}_{\ell}$ in our
problem, we start from the analogous of Eq.~\eqref{eq:def_phi_y}
\begin{equation}
    Y^{(\pm)}_{\ell}(\rtl) = X(\rt(\rtl)) / \sqrt{\dd \rt/\dd \rtl}.
\end{equation}
Notice that
\begin{equation}
    \frac{\dd \rt}{\dd \rtl} = \frac{\dd \rt}{\dd r} \frac{\dd r}{\dd \rtl} = \frac{1}{a} \, (a \cs) = \cs,
\end{equation}
where, recall, we defined $a = N f$.

\begin{figure}[t]
    \includegraphics[width=1\columnwidth]{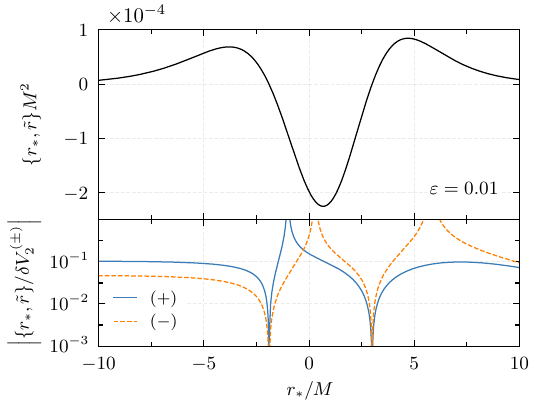}
    \caption{
    The Schwarzian derivative and its comparison to the EFT corrections to
    the gravitational potential. Top panel: the contribution of the Schwarzian
    to the potential, shown in units of $M$, as a function of $r_*$
    for $\varepsilon = 0.01$.
    Bottom panel: The absolute value of the ratio between the Schwarzian
    derivative's contribution and the EFT correction, denoted as $\delta
    V_{\ell}^{(\pm)}$, for the $\ell = 2$ multipole. The blue curve represents
    the comparison for the polar potential ($+$), while the orange curve
    corresponds to the axial potential ($-$).
    The Schwarzian contribution is subleading compared to $\delta
    V_{\ell}^{(\pm)}$ except when the latter function changes sign leading to
    singular points in the plot.
    }
    \label{fig:schw_plot}
\end{figure}

In Fig.~\ref{fig:schw_plot}, we show in the top panel the Schwarzian $\{\rt,
\rtl\}$ as a function of $r_{*}$. Its shape is not affected by $ \varepsilon $,
which gives the overall scaling.
In the bottom panel, we compare the Schwarzian with the $c^2_{\rm s} \,
V_{\ell}^{(\pm)}$, which we define as $\delta V_{\ell}^{(\pm)} $, for $\ell =
2$. This shows that the Schwarzian is typically smaller by an order of
magnitude compared to $\delta V_{\ell}^{(\pm)}$, except around points where the
latter crosses zero.

As a final remark, we connect our result to the ``parametrized master equation'' of Ref.~\cite{Cardoso:2019mqo}, which models gravitational modifications through simple power-law corrections to the potential. The master equation for $Y_\ell^{(\pm)}$ with potential $W_{\ell}^{(\pm)}$ [Eq.~\eqref{eq:W_pot}] fits directly into this formalism. Specifically, it matches the structure of their Eq.~(A.4) [notice that the Schwarzian derivative~\eqref{eq:schw_fun_r} vanishes at the horizon], which in turn ensures its equivalence to their main Eq.~(1) through a rescaling of $Y_\ell^{(\pm)}$.

\subsection{Numerical simulations} \label{app:schwarz:simulation}

We now study the time-domain
evolution of a Gaussian pulse impinging upon the ``Schwarzian potential.''
That is, we evolve Eq.~\eqref{eq:eqs_rwz_schwarz},
\begin{equation*}
    [ -\partial_{tt} + \partial_{\rtl\rtl} - W^{(\pm)}_{\ell}(r) ] \,
    Y^{(\pm)}_{\ell}(t,r) =
    0,
\end{equation*}
omitting the potential $V^{(\pm)}_{\ell}$ in
Eq.~\eqref{eq:W_pot}, i.e.,
\begin{equation} \label{eq:W_scharz}
    W^{(\pm)}_{\ell} = - \{\rt, \rtl\}.
\end{equation}
This set up is the ``Schwarzian-frame'' dual to the case~\ref{itm:case_c},
in which,
\begin{equation}
    \css \neq 1, \quad \textrm{and} \quad V^{(\pm)}_{\ell} = 0,
\end{equation}
we discussed in Sec.~\ref{sec:simulations:eft}.
The initial data is similar to
Eqs.~\eqref{eq:id_static} to \eqref{eq:id_params}, except that we evolve the
variable $Y^{(\pm)}_{\ell}$ (not $X^{(\pm)}_{\ell}$), and we use $\tilde{r}$
(not $r_{\ast}$) for the spatial coordinate.
For example, the pulse starts at $\tilde{r}^{\,\rm med} = 100M$.

\begin{figure}[t]
    \includegraphics[width=1\columnwidth]{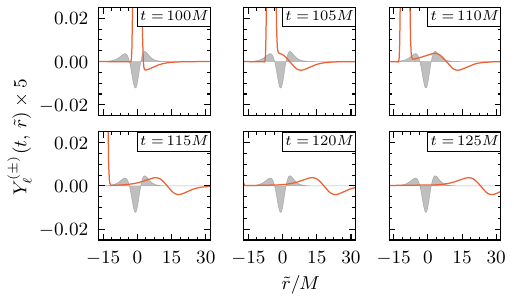}
    \caption{Snapshots of the incident perturbation (solid line and
    rescaled by a factor of 5) as it impinges with constant propagation
    speed into the ``Schwarzian potential,'' for $\varepsilon = 0.05$,
    represented by the filled gray curve.
    As the left-moving perturbation reaches the potential,
    part of it is reflected, as shown in the top panels. This reflected
    pulse propagates rightward (see bottom panels) and is measured
    by an observer at $\tilde{r}^{\,\rm ext}$ as the waveform shown in
    Fig.~\ref{fig:mystery}.}
    \label{fig:snapshots_schwarzian}
\end{figure}

Because our initial data is momentarily static and located in a region where
Eq.~\eqref{eq:W_scharz} is negligible, the Gaussian splits symmetrically into
left and right-moving parts.
Figure~\ref{fig:snapshots_schwarzian} shows snapshots
of the left-moving component (solid line) as it interacts
with the potential~\eqref{eq:W_scharz} (color-filled region).
The dynamics unfolds similarly to what we found in Fig.~\ref{fig:snapshots_variable_speed},
when the pulse entered the region where its propagation velocity is position
dependent.

\begin{figure}[t]
  \includegraphics[width=\columnwidth]{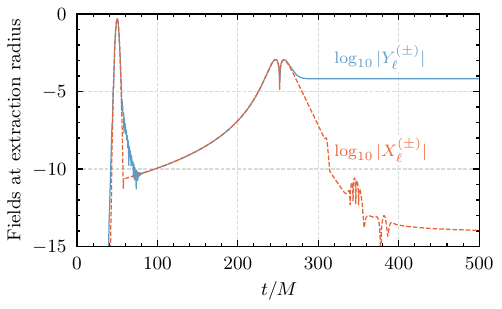}
  \caption{
      The fields $X^{(\pm)}_{\ell}$ (dashed line) and
      $Y^{(\pm)}_{\ell}$ (solid line) extracted at $\tilde{r}^{\,\rm ext} =
      r_{*}^{\rm ext} = 150M$, for similar momentarily static Gaussian initial
      data centered at $\tilde{r}^{\,\rm med} = r_{*}^{\rm med} = 100M$.
      The initial data splits symmetrically into left and right-moving
      components; the latter reaches the extraction radius around
      $t = 50M$.
      The peaks around $t=250M$ are the part of the left-moving component that was
      reflected after interacting with the region of variable propagation
      speed, for $X^{(\pm)}_{\ell}$, or the Schwarzian potential, for
      $Y^{(\pm)}_{\ell}$.
      The waveforms are similar, but at later times: while
      $X^{(\pm)}_{\ell}$ decays, $Y^{(\pm)}_{\ell}$ becomes constant.}
\label{fig:mystery}
\end{figure}

In Fig.~\ref{fig:mystery}, we show the waveforms
\begin{equation}
Y^{(\pm)}_{\ell}(t,\tilde{r}^{\,\rm ext})
\quad \textrm{and} \quad
X^{(\pm)}_{\ell}(t,r_{*}^{\rm ext}),
\end{equation}
extracted at $\tilde{r}^{\,\rm ext} = r_{*}^{\rm ext} = 150M$.
Starting from $t=0$, we see two pulses reaching the extraction radius.
The first peak, around $t=50M$, is the right-moving component reaching the
extraction radius.
The second peak, around $t=250M$, is the part of the left-moving component that
was then reflected after interacting with the region of variable propagation
speed, for $X^{(\pm)}_{\ell}$, or the Schwarzian potential, for
$Y^{(\pm)}_{\ell}$. See Figs.~\ref{fig:snapshots_variable_speed}
and~\ref{fig:snapshots_schwarzian}, respectively.
The two waveforms are similar, except for a notable difference at later times:
while $X^{(\pm)}_{\ell}$ decays, $Y^{(\pm)}_{\ell}$ approaches an approximately
constant value $6 \times 10^{-5}$. This value is around 5 orders of magnitudes
larger than the numerical errors in our code.
(Take as reference the waves reflected at the grid boundaries in the
$X^{(\pm)}_{\ell}$ curve around $t \approx 350M$.)
We found that this ``late-time constancy'' is related to the field
$Y^{(\pm)}_{\ell}$ settling to a nearly constant spatial profile at the end
of our simulation.
This happens as long as $\varepsilon$ is nonzero: the smaller $\varepsilon$ is,
the smaller the constant value the field settles to.
We did an extensive study to understand the origin of this behavior,
but ended empty handed. For this reason, we used Eq.~\eqref{eq:eqs_rwz} for our
simulations instead Eq.~\eqref{eq:eqs_rwz_schwarz}.

It would be interesting to see if there is a relation between our results
and the ``anomalous permanent displacement'' of the d'Alembert solution studied
in Ref.~\cite{Kroon:2024bnr}.

\section{Dependence of the quasinormal frequency extraction on the initial data} \label{app:qnm_fit_med_var}

Throughout the main text, we used a momentarily static Gaussian initial data in
our simulations; see Eqs.~\eqref{eq:id_static} to~\eqref{eq:id_params}.
In this appendix, we summarize the outcome of additional numerical
experiments where we varied the median of the Gaussian
from the value $r_{*}^{\rm med} = 100M$, used in the main text, down to $30M$, while keeping
the amplitude $A = 1$ and width $\sqrt{2} \, \sigma = 1.5 M$ fixed.
This brings the initial data from a region where it has negligible spatial
overlap with the effective potential $V_{\ell}^{(\pm)}$ closer to the
potential's peak; see Fig.~\ref{fig:potentials}.

\begin{figure}[t]
  \includegraphics[width=1\columnwidth]{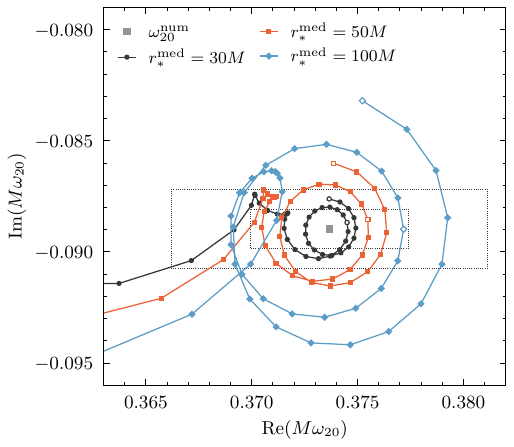}
  \caption{Trajectories in the complex plane for the best-fit
      quasinormal frequencies $\omega_{20}$ in general relativity as $t_0 = t - \tp$
      varies from zero to $30M$
      (first unfilled marker following the trajectories left to right) and
      $50M$ (second unfilled marker). The rectangles represent
      values $\pm1\%$ and $\pm2\%$ away from reference value of $\omega_{20}$ [filled square; cf.~Eq.~\eqref{eq:qnm_20}.]
     By decreasing $r_{*}^{\rm med}$, we can better recover $\omega_{20}$ and also
maintain less than $2\%$ error for a wider range of values of $t_0$.}
  \label{fig:fit_gr_var_med}
\end{figure}

We carried out three simulations with $r_{*}^{\rm med} = 30M$, $50M$, and
$100M$, and then examined the waveforms' spectral content using the method
we described in Sec.~\ref{sec:simulations:eft:qnm_fit}.
We found that we could extract more accurately the fundamental quasinormal
frequency from our time-domain data as $r_{*}^{\rm med}$ decreases.
Figure~\ref{fig:fit_gr_var_med} summarizes our results for the extraction of
$\omega_{20}$ whose numerical value in general relativity is given by
Eq.~\eqref{eq:qnm_20}.
As in Fig.~\ref{fig:l2n0_fits} we show the trajectories in the complex plane
for the best-fit values of $\omega_{20}$ as we vary the start of the fit time
window $t_0$ from zero up to $50M$ in steps of $1M$. We indicate with an unfilled
marker the values obtained for $t_0 = 30M$ and $t_0 = 50M$.
The three trajectories represent the three values of $r_{*}^{\rm med}$,
as indicated in the figure's legend.
We see that when $r_{*}^{\rm med} = 30M$, the curve spirals closer to the true
$\omega_{20}$ value (solid square) in comparison to the $r_{*}^{\rm med} = 50M$
and $100M$ cases.
We also find that even when we extend $t_0$ to $50M$, the recovered values of
$\omega_{20}$ still remain within 2\% error to the true value. This does not
occur in the other two cases.

\begin{figure}[t]
  \includegraphics[width=1\columnwidth]{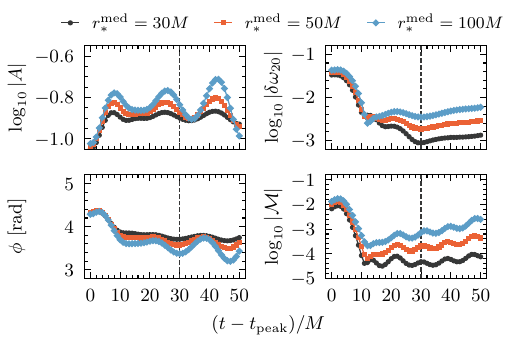}
  \caption{Dependence of the mismatch and the best-fit parameters in our fitting
      model~\eqref{eq:qnm_fit} as a function of $t_0 = t - \tp$. We see the best-fit
      parameters are more stable when varying $t_0$ and have smaller
      mismatches by making the peak of the momentarily-static Gaussian initial
      data, $r_{*}^{\rm med}$, closer to the peak of the effective potential.}
  \label{fig:fit_gr_coefs_var_med}
\end{figure}

In Fig.~\ref{fig:fit_gr_coefs_var_med}, we show the dependence of the best-fit
parameters of our model~\eqref{eq:qnm_fit} as a function of $t_0 = t - \tp$.
The vertical lines at $t_0 = 30M$ indicate the largest value of $t_0$ used in the main
text.
We see that the case $r_{*}^{\rm med} = 30M$ displays less variability in its
amplitude $A$ with respect to $t_0$ (as long as $t_0 \gtrsim 10M$) among the
three cases. It is also the case that reaches the smallest absolute error,
$\delta \omega_{20}$, has the smallest mismatch, ${\mathcal M}$.
We were unable to identify the reason for these results. We speculate that
bringing the Gaussian closer to the potential causes the quasinormal-mode
amplitude of the fundamental frequency to be larger than that of its overtones. This
would make the ringdown waveform become more ``fundamental-mode dominated'' and,
consequently, our single-mode fits more accurate and stable.
Our study in this appendix was done in general relativity, but we reached
similar conclusions in the EFT.

\bibliography{biblio}
\end{document}